\documentclass[prc,aps,nofootinbib,showkeys,showpacs,twocolumn]{revtex4}
\usepackage{epsfig}
\usepackage{graphicx}
\usepackage{amssymb}
\usepackage{color}
\usepackage{amsmath}

\graphicspath{{figures/}}

\begin{document}

\title{Microscopic description of pair transfer between two superfluid Fermi systems: combining phase-space averaging and combinatorial techniques  
}
  
\author{David Regnier} \email{regnier@ipno.in2p3.fr}
\affiliation{Institut de Physique Nucl\'eaire, IN2P3-CNRS, Universit\'e Paris-Sud, Universit\'e Paris-Saclay, F-91406 Orsay Cedex, France}

\author{Denis Lacroix} \email{lacroix@ipno.in2p3.fr}
\affiliation{Institut de Physique Nucl\'eaire, IN2P3-CNRS, Universit\'e Paris-Sud, Universit\'e Paris-Saclay, F-91406 Orsay Cedex, France}

\author{Guillaume Scamps} 
\affiliation{Center for Computational Sciences, University of Tsukuba, Tsukuba 305-8571, Japan}

\author{Yukio Hashimoto}
\affiliation{Center for Computational Sciences, University of Tsukuba, Tsukuba 305-8571, Japan}

 \date{\today}
\begin{abstract} 
In a mean-field description of superfluidity, particle number and gauge angle are treated as quasi-classical conjugated variables. 
This level of description was recently used to describe nuclear reactions around the Coulomb barrier. Important effects of the relative gauge angle between two identical
superfluid nuclei (symmetric collisions) on transfer probabilities and fusion barrier have been uncovered. A theory making contact with experiments should at least average over different initial relative gauge-angles.
In the present work,
we propose a new approach to obtain the multiple pair transfer probabilities between superfluid systems. This method, called Phase-Space combinatorial (PSC) technique, relies both on phase-space averaging and combinatorial arguments to infer the full pair transfer probability distribution at the cost of multiple mean-field calculations only. After benchmarking this approach in a schematic model, we apply it to the collision $^{20}$O+$^{20}$O at various energies below the Coulomb barrier. The predictions for one pair transfer are similar to results obtained with an approximated projection method whereas significant differences are found for two pairs transfer. Finally, we investigated the applicability of the PSC method to the contact between non-identical superfluid systems. 
A generalization of the method is proposed and applied to the schematic model showing that the pair transfer probabilities are reasonably  reproduced.  The applicability of the PSC method to asymmetric nuclear collisions is investigated for the $^{14}$O+$^{20}$O collision and it turns out that unrealistically small single- and multiple-pair transfer probabilities are obtained.
This is explained by the fact that relative gauge angle play in this case a minor role in the particle transfer process compared to other mechanisms such as equilibration of the charge/mass ratio. We conclude that the best ground for probing gauge-angle effects in nuclear reaction and/or for applying the proposed PSC approach on pair transfer is the collisions of identical open-shell spherical nuclei. 
\end{abstract}

  
\maketitle

\section{Introduction}


Although its contribution to the energy is rather small, superfluidity plays a significant role  in static properties of atomic nuclei \cite{Bro13,Bri05}. 
Its influence on time evolution of nuclei is however scarcely known.  One effect that could be naturally anticipated is the enhancement of simultaneous transfer of two nucleons when the two particles form a pair \cite{Von01-b,Gra12}. Another predicted effect is the global increase of pair transfer when coherent oscillations of pairs exist, the so-called pairing vibration \cite{Bes66,Bro73,Gam12}. While a possible effect of pairing on transfer is rarely contested \cite{Von01-b,Gra12}, its quantitative influence remains to be clarified. A detailed analysis has been for instance made recently in Ref.~\cite{Pot17} where different levels of description from a pure mean-field to beyond mean-field have been considered.  In parallel, to understand the competition between transfer and fusion reaction, new highly accurate experimental measurements of transfer probabilities have been achieved  
giving test-bench for nuclear models \cite{Cor11,Mon14,Mon16,Raf16,Sca16}. 
 
In recent years, intensive efforts have been made to include pairing into dynamical microscopic theories \cite{Ave08,Eba10,Ste11,Has12,Has13,Sca13} in order to simulate both isolated nuclei and/or collisions between nuclei. These approaches offer the possibility to understand the influence of superfluidity on the dynamics from a different point of view than traditional approaches where nuclear structure is treated separately from nuclear reaction.  The basic tools to include superfluidity is the Hartree-Fock-Bogoliubov (HFB) or 
BCS theory where the $U(1)$ symmetry associated to the gauge angle is spontaneously broken to include pairing correlations in a simple way. 
Similarly to interacting bosonic systems it is quite natural to investigate if the interaction of two nuclei is affected by their relative gauge angles when they enter into contact. Recently, several works have uncovered rather large effects of the relative gauge angle between 
identical nuclei. 
This case of symmetric collisions is quite special in the sense that effects such as charge/mass equilibration between collision partners are absent. This context allows the pairing fluctuations to become a major driver of the pair transfer. 
A first hint on the role of gauge angle in symmetric reactions was given in Ref.~\cite{Has16} where its influence on pair transfer has been addressed for the first time within TDHFB. It was indeed found in particular that particle transfer is sensibly affected by the relative gauge angle between nuclei. 
The sensitivity to the gauge angle has been further explored in Ref.~\cite{Sek17,Mag17,Bul17} confirming again its importance in the particle transfer process. Besides, these works emphasize two surprising behaviors. First, when fusion does not occur, the kinetic energy of fragments after re-separation is significantly affected by the initial relative phase in gauge space. Second, the relative angle changes the fusion threshold leading to a considerable contribution to the so-called extra-push energy.

The fact that some spontaneous symmetry breaking affects the physics close to the Coulomb barrier is by itself not a surprise. Another typical example is given by the spontaneous breaking of the rotational symmetry at the mean field level that leads to deformed nuclei in their intrinsic frame. The role of the gauge angle is in this case replaced by the relative orientation between the two main axis of deformation of the collision partners. These collective degrees of freedom lead to fluctuations in the Coulomb barrier properties that can be probed experimentally~\cite{Das98,Row91}. Note that an extensive discussion of the possible role of spontaneous symmetry breaking (\textit{e.g.} deformation in real and/or gauge space) on transfer reactions can also be found in Ref.~\cite{Bro91}.
Mean field theories like Hartree-Fock and/or Hartree-Fock Bogoliubov provide a proper framework to describe spontaneous symmetry breaking in the sense that these approaches naturally yield deformed ground states. However, they treat the 
collective variables classically while, in principle, a fully quantum collective approach would be required. To overcome this difficulty, the mean-field picture requires some mechanism to restore the broken symmetries of the total system before, during and after the collisions. 
A natural technique to insure proper symmetries in state of the art nuclear structure studies is to project the trial wave-function on good particle number, parity, angular momentum \cite{Rin81, Ben03, Egi16}. This technique is however much more involved for nuclear reactions where two evolving systems with various exit channels are considered.  Up to now, the projection technique has been essentially used after the reaction involving either two normal~\cite{Sim10} or 
only one superfluid nucleus \cite{Sca13}.

A first attempt was made recently to use projection technique during the evolution~\cite{Sca17,Sca17b}. 
An important advantage is eventually to be able to treat interferences between different mean-field trajectories. The technique proposed in~\cite{Sca17} turns out to be rather involved even if a drastic approximation was made on the mixing of different 
mean-field trajectories.  In addition, two difficulties show up: (i) the results strongly depend on the conventions for the equation of motion used to solve the TDHFB equations (ii) the final transfer probabilities might have rather important oscillations depending on the phase evolution convention even when the two nuclei are well separated.

Based on the recent investigation on the role of gauge angle on nuclear reactions, there is a number of emerging interrogations: 
the first one, is that the gauge angle itself is a concept that only has a meaning in a symmetry breaking theory while the number of protons and neutrons 
in nuclei are fixed. Then, are the predicted effects surviving in a particle number conserving theory ? If yes, do the huge effect predicted on reactions in Ref.~\cite{Mag17} persists once the symmetry is properly restored ? 
Are these effects still important in collisions involving a target different from the projectile ?
Behind these questions, one may wonder if the classical mechanics mean-field approach based on gauge angle has a physical observable reality in nuclei ? Another difficulty is that contrary to cold atoms, and even if nuclei might have rather large number of nucleons, the true number of nucleons (essentially those close to the Fermi energy) forming pairs is rather small. Then, finite size corrections to the BCS/HFB approaches are expected to be significant as well as correlations much beyond the mean-field picture.

This article is organized as follows. First, we discuss some of the points raised above and question if a classical picture for the gauge space  is meaningful in the context of transfer reactions (section~\ref{sec:description}). In section~\ref{sec:beyondmf} we review several existing technics to restore the symmetry associated to the number of particles in the context of nuclear reactions and propose a new method that we call Phase-Space Combinatorial (PSC) technique. This method is benchmarked in a schematic model of collision between two identical systems (symmetric collisions) and at energies below the Coulomb barrier. The section~\ref{sec:o20o20} highlight its application to the realistic collision $^{20}$O + $^{20}$O. Finally, we discuss in section~\ref{sec:gen_asym} the applicability of the PSC method to the case of asymmetric collisions, i.e. when the two superfluids are different,   
for which the main driver of particle transfer may not be the residual pairing interaction.

\section{Description of transfer reactions in a schematic model}
\label{sec:description}

In this section, we focus on the exchange of pairs of fermions happening two symmetric and superfluid systems come to contact. We follow Ref. \cite{Sca17,Sca17b} and consider a minimal schematic Hamiltonian to describe the transfer between a superfluid system $A$ and a superfluid system $B$.
In this simple model, the systems are assumed to stay at any time in fully paired states.
The Hamiltonian is a sum of three terms
\begin{eqnarray}
 \label{eq:hab}  
H & = & H_A + H_B + V(t) . \label{eq:htot}
\end{eqnarray}
The operator $H_A$ (resp. $H_B$) describes the isolated Fermi system $A$ (resp. $B$) and is supposed to take the form of a simple pairing Hamiltonian (see for instance \cite{Bri05}). 
\begin{eqnarray}
H_A&=& \sum_{k>0}^{\Omega_A} \varepsilon^A_k (a^\dagger_k a_k + a^\dagger_{\bar k} a_{\bar k}) 
- g_A \sum_{k\neq l>0}^{\Omega_A}  a^\dagger_{ k} a^\dagger_{\bar k} a_{\bar l} a_l, \label{eq:ha} \nonumber \\
H_B&=& \sum_{k>0}^{\Omega_B} \varepsilon^B_k (b^\dagger_k b_k + b^\dagger_{\bar k} b_{\bar k}) 
- g_B \sum_{k\neq l>0}^{\Omega_B}  b^\dagger_{ k} b^\dagger_{\bar k} b_{\bar l} b_l .  \label{eq:hb} 
\end{eqnarray}
Here, $\{a_k, a_{\bar k}\}$ (resp. $\{b_k, b_{\bar k}\}$) correspond to a set of $\Omega_A$ (resp. $\Omega_B$) pairs of states, where ${\bar k}$ denotes the time-reversed state of $k$. 
The term $V(t)$ describes the interaction between  the two systems during the contact time. 
As stated in Ref.~\cite{Die70}, two mechanisms may drive the transfer of particles between two colliding nuclei. The first one is the tunneling of single particles coming from the mean field and can be mimicked by operators $a_k^\dagger b_l$ whereas the second one is the direct transfer of pairs due to the residual pairing interaction and is related to $a^\dagger_{ k} a^\dagger_{\bar k} b_{\bar l} b_l$. In the present model, the first 
term is omitted and therefore it implicitly assumes that the transfer of pairs mainly occurs simultaneously. 
As we will see below, such a model can eventually model the pair transfer during symmetric heavy-ion collisions 
but not asymmetric reactions. In the present work, we mainly concentrate on symmetric collisions where large gauge-angle effects has been uncovered. Then, neglecting the single-particle tunneling, the transfer process is described using~\cite{Sca17b}:
\begin{eqnarray}
V(t) & = & v(t) \sum_{k>0}^{\Omega_A}\sum_{l>0}^{\Omega_B} \left(a^\dagger_{ k} a^\dagger_{\bar k} b_{\bar l} b_l +  b^\dagger_{ l} b^\dagger_{\bar l}  a_{\bar k} a_k \right). 
\label{eq:vab}
\end{eqnarray}
The time dependency of the coupling is assumed to be Gaussian and centered at the collision time $t=0$.
\begin{equation}
 v(t) = v_0 \operatorname{exp}\left[ -t^2/\tau_c^2 \right],
 \end{equation}
where $\tau_c$ is the interaction time.
Under simple assumptions, the two parameters of the coupling strength $v(t)$ may be related to a few characteristics of the input channel namely the charges $Z_A,Z_B$, the reduced mass of the colliding nuclei, the relative kinetic energy and  the scattering angle in the center of mass frame~\cite{Die70}. 

In the following, for the sake of compactness, we denote generically by $(c^\dagger_k, c^\dagger_{\bar k})$ a pair of states belonging to either 
the system $A$ or $B$. Then the total Hamiltonian simply writes:
 \begin{eqnarray}
H & = & \sum_{k>0} \varepsilon_k (c^\dagger_k c_k + c^\dagger_{\bar k} c_{\bar k}) 
- \sum_{k\neq l>0}  G_{kl}(t) c^\dagger_{ k} c^\dagger_{\bar k} c_{\bar l} c_l. \label{eq:hab_bis}
\end{eqnarray} 
With these notations, $\varepsilon_k = \varepsilon^A_k$ (resp. $\varepsilon^B_k$) if ($k$, $\bar k$) belong to $A$ (resp. $B$). The matrix element $G_{kl}$ equals respectively to $g_A$, $g_B$ and $v(t)$ 
if the couple of indices $(k,l)$ refers to states that are both in $A$, both in $B$, or one in $A$ and one in $B$.  

A direct diagonalization technique applied in every subspace of given seniority gives the complete set of eigenstates of the pairing Hamiltonian~\cite{Vol01}. This direct technique is possible as long as the single-particle Hilbert space is not too large. 
This advantageous feature of the pairing Hamiltonian has been widely leveraged to study the static properties of a variety of small superfluid systems~\cite{Von01,Bri05,Dea03} as well as to test approximate treatment of pairing~\cite{Duk04}.
The full Hamiltonian (\ref{eq:hab}) was first proposed to study transfer reaction in Ref.~\cite{Die70}
and further analyzed in \cite{Die71,Har71}. It was also used by Broglia~\textit{et. al.} to discuss the semi-classical nature of the mean-field approximation (see for instance \cite{Bro91}). 
The exact solution of this schematic Hamiltonian guided us to propose, in this paper, an approximate treatment of the pair transfer between two colliding nuclei. This approximate treatment aims at being applicable to heavy systems where the direct diagonalization of the pairing Hamiltonian becomes impractical. We first discuss below briefly the exact and mean-field solutions. 
  
\subsection{Exact solution}
\label{sec:exasol}

We assume at initial time that each subsystem is in its ground state. In these ground states, all the particles are paired. Since the complete Hamiltonian (\ref{eq:htot}) does not break pairs during the evolution, the total wave-function of the composite system remains in a subspace where all nucleons are paired during the whole evolution.
Calculations can therefore be performed in the basis of orthonormal states $\{ |n\rangle \}$ defined as:
\begin{align}
 |n\rangle = \prod_{k>0}^{\Omega} (c^\dagger_k c^\dagger_{\bar{k}})^{n_k} |0\rangle, \quad n_k=0,1
\end{align}
where $\Omega = \Omega_A+\Omega_B$. Due to the invariance of the total particle number respected by the model Hamiltonian, this basis can even be reduced to states verifying the condition
\begin{equation}
 \sum_{k>0}^{\Omega} n_k = \frac{1}{2} (N_A^0 + N_B^0),
\end{equation}
where $N^0_A$ and $N^0_B$ denote the initial numbers of particle in system $A$ and $B$ respectively.
The string of bits $n_\Omega n_{\Omega-1} \cdots n_1$ totally defines the state $|n\rangle$ and provides a direct mapping between this state and the integer $n$ having $n_\Omega n_{\Omega-1} \cdots n_1$ for binary representation. The initial state of each subsystem is determined by diagonalizing its own Hamiltonian within the subspace spanned by $|n\rangle$ states with appropriate number of particles, i.e. $ \sum_{k\in A} n_k = N^0_A/2$ and $\sum_{k\in B } n_k = N^0_B/2$. 
The time-evolution of the total wave-function $| \Psi \rangle$ is then obtained by solving the coupled-channel equations on the components 
$c_n(t)$ with:
\begin{eqnarray}
| \Psi (t) \rangle &=& \sum_n c_n(t)  |n\rangle. \label{eq:psiexact} 
\end{eqnarray}  
From this coefficients, any quantity related to transfer can be computed.  For instance, the probability to have $(N_A, N_B)$ particles at final time  $t_\infty$ is given by:
\begin{align}
\label{eq:pfinal}
P(N_A,N_B, t_\infty)  =  \sum_{n\in \mathcal{E}(N_A,N_B)} |c_n(t_\infty)|^2,
\end{align}
where the sum runs over the set of states $\mathcal{E}(N_A,N_B)$ defined by the condition:
\begin{equation}
 \sum_{k\in A} n_k = N_A/2, \quad \sum_{k \in B} n_k = N_B/2.
\end{equation}
Since the total particle number $N$ is conserved, the final probability given by (\ref{eq:pfinal}) is zero if $N_B \neq N - N_A$. 
In the following, we introduce the exact pair transfer probability $P^{exa}_{xn}$ that is equal to
\begin{equation}
\label{eq:pexa}
P^{exa}_{xn} = P(N_A= N^0_A+x, N_B=N_B^0-x, t_\infty).
\end{equation}
We follow the standard terminology and call addition (resp. removal) probabilities, 
the probability for $x>0$ (resp. $x < 0$), implicitly assuming that the addition/removal is defined with respect to the system $A$. 

Using this practical scheme, we compute the multiple pair transfer probabilities during the collision of two identical systems. This example is used as a reference calculation throughout this article. Each system consists of one degenerate shell with single particle energies set to $\epsilon_k^A=\epsilon_k^B=0$. The shell degeneracy is set to  $\Omega_A=\Omega_B=6$ and both systems are initialized in their half filling situation, \textit{i.e.} $N_A^0/2=N_B^0/2 = 3$ pairs of particles. This simple assumption can be regarded as a minimal description of the transfer of nucleon pairs 
from one degenerate shell of nucleus $A$ (resp. $B$) to another degenerate shell of nucleus $B$ (resp. $A$). 
The pairing strength is assumed to be the same in the two systems $g=g_A=g_B=1$ MeV.
The characteristic contact time defining the coupling Hamiltonian is set to $ \tau_c =  0.28 \,\hbar /g$ and the evolution takes place from $t_0 = -2.28$ to $t_\infty= + 2.28$ in the same units. This time interval is wide enough to probe the asymptotic regime both before and after the collision.
The initial product state is evolved in time using a Taylor expansion of the exponential propagator up to fourth order with a sufficiently small time step $dt = 10^{-4}\, \hbar/g$. After the collision, the transfer probabilities are recovered using Eq. (\ref{eq:pfinal}). 

\begin{figure}[htbp]
\includegraphics[width=0.9\linewidth]{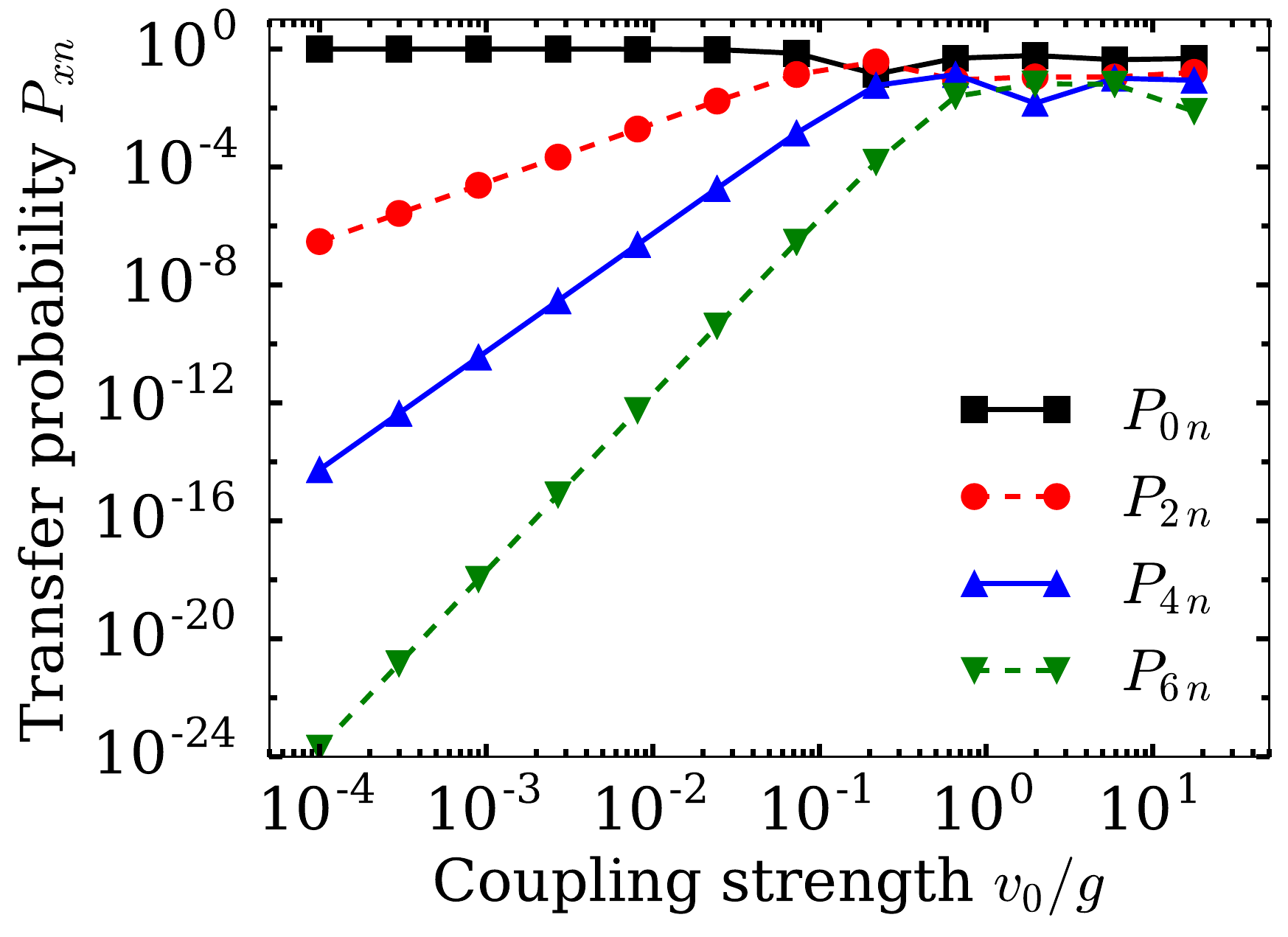}
\caption{Exact asymptotic probabilities of multiple pair transfer as a function of the coupling strength $v_0/g$ obtained for 
the symmetric degenerate case with $\Omega_A=\Omega_B=6$ and $N_A^0=N_B^0=6$.}
\label{fig:pv0_exa} 
\end{figure}
The figure~\ref{fig:pv0_exa} shows the results obtained for a wide range of coupling strength $v_0/g$. Two regions of coupling strength can be identified. 
For $v_0/g > 2.10^{-2}$, the probabilities to transfer several pairs have the same order of magnitude. 
This region corresponds to a highly non-perturbative regime where strong quantum interferences between different transfer channels 
play an important role. 
On the other hand, for $v_0/g < 2.10^{-2}$, the interaction acts as a small perturbation. 
The probability to transfer one pair becomes prominent compared to multi-pair transfer. 
The perturbative nature of the transfer can be directly inferred from the simple scaling behavior 
of $P_{xn}$ observed in Fig.~\ref{fig:pv0_exa}. Indeed, for small values of the coupling, 
$P_{xn}$ is proportional to $(v_0)^{x}$ which is consistent with the scaling deduced from the first 
non-zero term appearing in time-dependent perturbation theory (see appendix \ref{ap:exa}). It is finally worth mentioning that the 
observed probability dependence with $v_0/g$  looks very much the same as the observed evolution 
of transfer probabilities below the Coulomb barrier when plotted as a function of the minimal distance 
of approach \cite{Cor11,Mon14,Mon16,Raf16}. This underlines that the perturbative regime is certainly the 
most relevant  for these experiments.

\subsection{TDHFB solution}
\label{sec:tdhfb}

We now consider a mean-field description of transfer. 
The natural approach to circumvent the combinatorial growth of the exact Hilbert space is to restrain the system to a TDHFB ansatz. 
In the case of our model Hamiltonian, the HFB trial function reduces to a BCS one which takes the form
\begin{eqnarray}
| \Psi(t) \rangle & = & \prod_{k>0} (U^*_k(t)+ V^*_k(t) c^\dagger_k c^\dagger_{\bar k}) | 0\rangle .
\label{eq:bcssimple}
\end{eqnarray}
The single-particle occupation numbers $n_k$ and the anomalous density components $\kappa_k$ defined as 
\begin{eqnarray}
n_k(t) & = & \langle c^\dagger_k c_k\rangle = \langle c^\dagger_{\bar k} c_{\bar k} \rangle  = |V_k(t)|^2,  \nonumber \\
\kappa_k(t)  &=& \langle c_{\bar k}c_k \rangle = U_k(t) V^*_k(t) , \nonumber
\end{eqnarray}  
contain all the information on the system.

The mean-field trajectory fulfills the Ehrenfest theorem $i\hbar \partial_t\langle \hat O \rangle  = \langle [\hat O, \tilde H] \rangle$ for any operator $\hat{O}$ that are linear combinations of $\{ c^\dagger_k c_k$, $ c_{\bar k} c_k $, $c^\dagger_{ k} c_{\bar k}^\dagger \}$ operators 
(here $\tilde H$ just means that it might contain or not the constraint on particle number $-\lambda \hat{N}$). This leads to the set of equations of motion:
\begin{eqnarray}
\left\{ 
\begin{array}{l}
\displaystyle   i\hbar \frac{d n_{k}}{dt}  =  \Delta^*_k \kappa_k - \kappa^*_k \Delta_k  ,   \\
\\
\displaystyle i\hbar \frac{d \kappa_k }{dt}   =  2 \tilde \varepsilon_k \kappa_k  + \Delta_k (2 n_k -1) .
\end{array}\right. 
\label{eq:tdhfb}
\end{eqnarray}
The pairing gap is defined as $\Delta_{k}  =  \sum_l G_{kl} \kappa_l $ and the single particle energies write 
$\tilde \varepsilon_k = \varepsilon_k  -\lambda_k$.
While the equations (\ref{eq:tdhfb}) can be solved directly, it is common to use instead the equation of motion on the quasiparticle components $(U_k, V_k)$:
\begin{eqnarray}
i\hbar \frac{d}{dt}
\left( 
\begin{array}{c}
U_k(t)    \\
V_k(t)       
\end{array}\right)
= \left( 
\begin{array}{c c }
\tilde \varepsilon_k   -  \gamma_k    &  \Delta_k  \\
 \Delta^*_k     &   - \tilde \varepsilon_k - \gamma_k
\end{array}
\right) 
\left( 
\begin{array}{c}
U_k(t)    \\
V_k(t)       
\end{array}\right) . \label{eq:general}
\end{eqnarray}
As previously noted \cite{Blo76,Eba10}, the above set of equations are not unique and can be solved using an arbitrary factor $\gamma_k$. This arbitrary factor brings a change in the evolution of the global phase of the quasi-particle vacuum  $|\Psi(t)\rangle$ while conserving the equation of motion (\ref{eq:tdhfb}). At the mean-field level, any expectation value will be independent of this factor and therefore $\gamma_k$ may be chosen arbitrarily~\cite{Blo76,Eba10}. If one tries to go beyond mean-field (\textit{e.g.} using theories requiring calculation of overlaps between 
different quasi-particle vaccua), the results strongly depend on the choice of this global phase. The fact that no specific choice has yet been established on first principle argument renders the treatment of the interaction between two superfluids rather tricky~\cite{Sca17}.

Within the TDHFB approach, the initial state is a product of two quasi-particle ground states associated with $A$ and $B$. The breaking of $U(1)$ symmetry for these two states leaves us with an arbitrary relative initial gauge angle $\theta^0_{AB}$ between the two subsystems:
\begin{eqnarray}
\kappa_A (0) \kappa^*_B (0) &=&  |\kappa_A(0) | |\kappa_B(0) | e^{i\theta^0_{AB}} .
\label{eq:kappa0}
\end{eqnarray}
As demonstrated in several works~\cite{Bul90,Sca17}, the final result of the collision treated within TDHFB depends on the initial relative 
gauge-angle. 


To better grasp this effect, we consider as in section~\ref{sec:exasol} the case of two symmetric fully degenerated systems.
In this simple model, the gauge angle and relative number of particles play the role of classical conjugated variables that obey simple coupled equations of motion. Due to the degeneracy in each system, only four parameters describe the TDHFB evolution: $n_A$, $n_B$, $\kappa_A$  and $\kappa_B$. Then, equation~(\ref{eq:tdhfb}) reduces to: 
\begin{eqnarray}
i\hbar \frac{d n_{A}}{dt} & = & \Omega_B v(t) (\kappa^*_B  \kappa_A - \kappa^*_A \kappa_B) =\frac{\Omega_B}{\Omega_A} \frac{d n_{B}}{dt},
\label{eq:nab}    \\
i\hbar \frac{d \kappa_A}{dt}  & = & 2 \tilde \varepsilon_A \kappa_A  + (\Delta_A + \Delta_{AB}(t)) (2 n_A -1), \label{eq:kappa_A} \\
i\hbar \frac{d \kappa_B}{dt}  & = & 2 \tilde \varepsilon_B \kappa_B  + (\Delta_B + \Delta_{BA}(t)) (2 n_B -1) ,  \label{eq:kappa_B} 
\end{eqnarray}
with 
\begin{eqnarray}
\Delta_A = g_A \Omega_A \kappa_A, ~~~\Delta_B = g_B \Omega_B \kappa_B, \nonumber \\
\Delta_{AB} = v(t) \Omega_B \kappa_B, ~~~ \Delta_{BA} = v(t) \Omega_A  \kappa_A.
\end{eqnarray}
At any time $t$, we may define the time-dependent relative angle $\theta_{AB}(t)$ by $\kappa_A \kappa^*_B  =  |\kappa_A| |\kappa_B| e^{i\theta_{AB}}$. Finally, introducing the number of particles $N_A(t)= 2\Omega_A n_A(t)$ in the subsystem $A$ (resp. $N_B(t)= 2\Omega_B n_B(t)$ in $B$) yields the following evolutions for the average particle numbers:
\begin{eqnarray}
 \frac{d N_{A}}{dt} & = & 4 v(t) \frac{\Omega_A \Omega_B }{\hbar} |\kappa_A(t)| |\kappa_B(t)| \sin \left[ \theta_{AB}(t)\right] \label{eq:Naevol} \\
 &=& - \frac{d N_{B}}{dt} . \nonumber 
\end{eqnarray}  
This evolution is rather complicated as it depends explicitly on the anomalous
density of each system. However, in the weak coupling regime, one might neglect the coupling term in Eq.~(\ref{eq:kappa_A}) and (\ref{eq:kappa_B}). Then, the evolution of the anomalous densities in the two subsystems become independent from each other and can be integrated in time:
\begin{align}
 \kappa_{A / B}(t) \simeq \kappa_{A / B}^0 \exp
 \left( -2i \omega_{A/B} t \right),
\end{align}
with $\hbar \omega_{A / B} = \epsilon_{A / B} - \lambda + g_{A / B} \Omega_{A/B} (n_{A/B} -1/2)$.
The relative gauge orientation then rotates with a nearly constant frequency $\omega_{AB} = \omega_A - \omega_B $:
\begin{align}
 \theta_{AB}(t)  \simeq - 2 \omega_{AB} t + \theta^0_{AB} .
 \end{align}
Reporting in the anomalous densities 
evolutions and integrating in time, we finally obtain an expression of $N_A(t)$:
\begin{eqnarray}
N_A(t) & = & N_A^0 + S_v (t) \sin \theta^0_{AB} - A_v(t) \cos \theta^0_{AB},
\label{eq:na}
\end{eqnarray}
 with 
 \begin{eqnarray}
S_v(t) &=& \int_{t_0} ^{t} W(s) \cos(2\omega_{AB} s)ds, \nonumber \\
A_v(t) &=&  \int_{t_0} ^{t} W(s) \sin(2\omega_{AB} s)ds, \nonumber
\end{eqnarray} 
and where we have introduced the notation 
\begin{eqnarray}
\label{eq:wt}
W(t) = \frac{4}{\hbar} v(t) \Omega_A \Omega_B |\kappa^0_A| |\kappa^0_B| .
\end{eqnarray}
In the asymptotic regime after the collision, we may choose $t_\infty=-t_0$ so that only the symmetric term will contribute to pair transfer. In the symmetric case, $\omega_{AB}\simeq 0$ and $S_v$ can be integrated explicitly as:
\begin{eqnarray}
S_v(t)  =  2 \left[ \frac{v_0 \tau_c}{\hbar}\right] v_0  \Omega_A\Omega_B |\kappa^0_A| |\kappa^0_B|  \sqrt{\pi} \nonumber \\
\times \left[{\rm erf}(t/\tau_c)    -  {\rm erf}(t_0/\tau_c) \right],
\label{eq:sv}
\end{eqnarray}  
with the standard definition  ${\rm erf}(x) =  \displaystyle \frac{2}{\sqrt{\pi}} \int_0^t e^{-s^2} ds $. 
As a conclusion, the average number of particle transferred in the weak coupling conditions is found to depend on the sinus of the relative gauge angle with an amplitude proportional to the a-dimensional parameter $(v_0 \tau_c /\hbar)$. 
\begin{eqnarray}
\label{eq:nasym}
N_A^\infty \simeq N_A^0 + S_v (t_\infty) \sin \theta^0_{AB} .
\end{eqnarray}

\begin{figure}[htbp]
\includegraphics[width=0.83\linewidth]{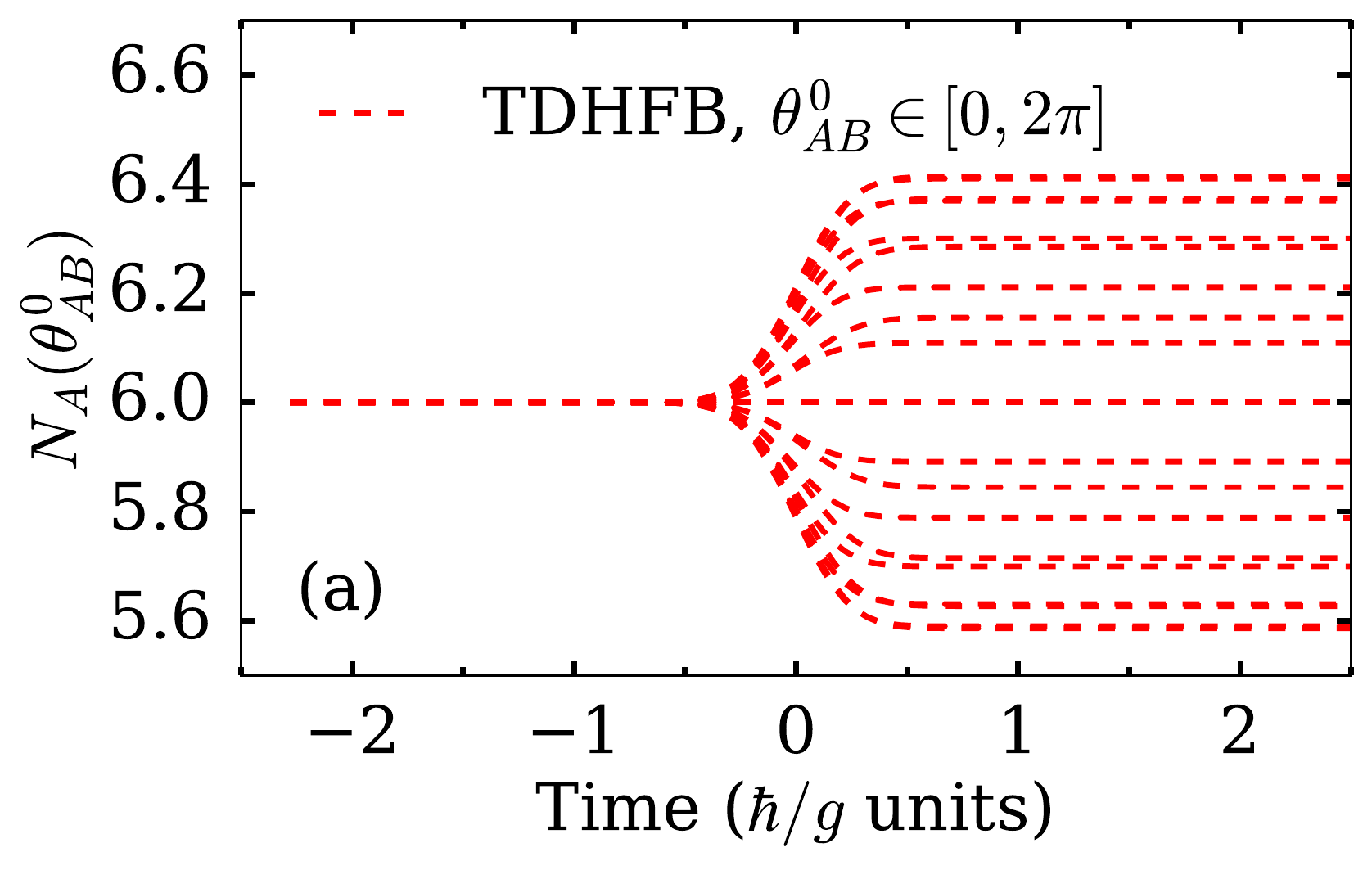}
\includegraphics[width=0.9\linewidth]{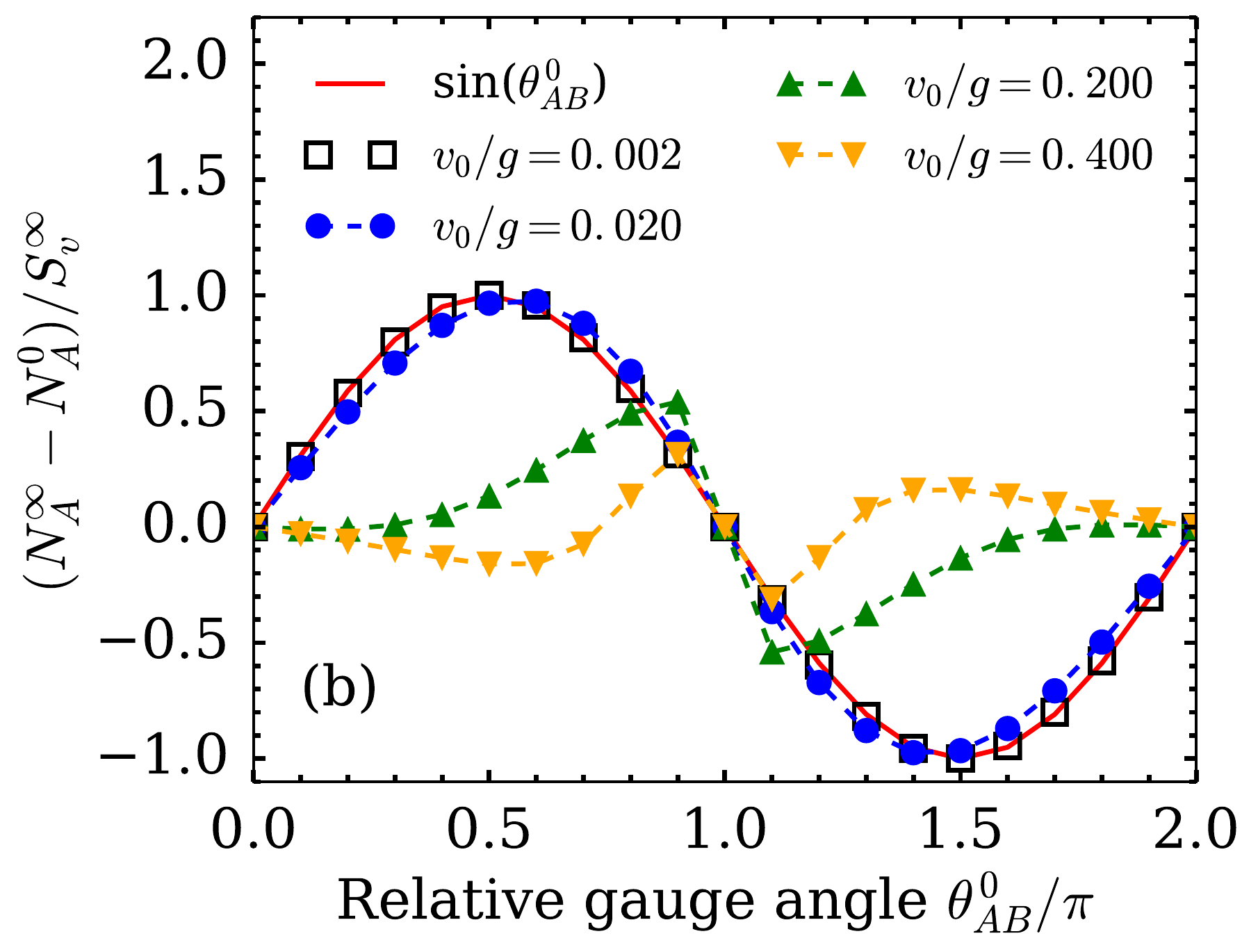}
\caption{(a) TDHFB evolutions of the particle number in the system $A$ starting from the relative gauge angles $\theta^0_{AB} = 2k\pi/20,\, 0\leq k < 20$ (same as Fig. 1 of Ref. \cite{Sca17} given here for completeness) and for a coupling strength $v_0/g=0.02$. (b) Asymptotic number of particles transferred to the system $A$ as a function of the initial relative gauge angle for four different coupling strengths. The number of transferred particles is normalized by $S_v^{\infty}$ and the weak coupling analytical formula, Eq. (\ref{eq:nasym}) is shown in red solid line. All TDHFB calculations are performed with a rescaling factor $\alpha=1.2$ on the interaction (cf. sec.~\ref{sec:crit}).}
\label{fig:phase1} 
\end{figure}

In Fig. \ref{fig:phase1}, the time evolution of $N_A$ obtained with mean-field is illustrated using different initial relative  phases for the case 
considered in sec.~\ref{sec:exasol} ($\Omega_A=\Omega_B=6$ and starting from the half filling configuration). 
In the top panel, we see that, starting from different $\theta^0_{AB}$, leads to different mean-field trajectories as soon as the two systems start to interact. Similar behavior has been obtained in ref. \cite{Sca17}. 
The bottom part of Fig. \ref{fig:phase1} shows the dependency of the asymptotic number of particle transferred as a function of the initial relative gauge angle. As we decrease the coupling strength, the asymptotic particle number tends toward the sinusoidal dependency given by Eq. (\ref{eq:nasym}). Already for $v_0/g =2.10^{-2}$, the agreement with the analytical formula is very good which is 
consistent with the beginning of the perturbative regime shown for the exact case in Fig.~\ref{fig:pv0_exa}.

\section{Description of pair transfer probabilities beyond mean-field}
\label{sec:beyondmf}

Mean field theories that treat superfluidity have inherent limitations. The first one is obviously the breaking of the $U(1)$ symmetry associated to particle number which makes difficult the extraction of transfer probabilities. Indeed, finite size effects and quantum fluctuations associated to the eventual symmetry restoration are anticipated to play a significant role, especially if the number of particles involved is small. A second limitation is the quasi-classical nature of the mean-field in collective space. 
{
In this section, we first give a non-exhaustive overview of some approaches that can include beyond mean-field effects as well as their possible shortcomings. We then present a new approach named phase-space combinatorial technique to compute pair transfer probabilities based on multiple TDHFB trajectories.

\subsection{Quantum methods to go beyond mean-field}

A natural extension to include quantum effects beyond mean field would be to use a path integral approach. In the simple degenerated model described here, it is indeed possible to include interferences between trajectories leading to the same final number of particles. From Fig. \ref{fig:phase1}, one indeed realizes that there are always at least two trajectories (weak coupling) or more (strong coupling) leading to the same $N^\infty_A$. Using the stationary phase approximation (SPA), we may estimate the transfer probability as a weighted average over different paths with the proper phase factor. This approach has already been applied with some success to transfer reaction, starting from the classical action $S_{\rm cl}(\theta_{AB}, N_A)$ and making additional efforts to access probabilities in the classically forbidden region~\cite{Bro91}.
We could have followed the same technique and most probably got reasonable probabilities. However, we anticipate that such a method may not be applicable for realistic nuclear collisions where the collective coordinates associated to the particle number and relative gauge angle are coupled to other degrees of freedom (\textit{e.g.} the relative distance between nuclei, the deformation, ...). A clear fingerprint of such a coupling is the large dependence of the fusion barrier with the gauge-angle empirically observed in Ref.~\cite{Mag17}. Treating this feature explicitly when performing the stationary phase approximation leads to an increase of the problem complexity that seems prohibitive. 
      
A second strategy consists in performing a proper quantum mixing of the TDHFB trajectories during the evolution. In such a framework, the symmetry is restored with an appropriate variation after projection (VAP) onto good particle number~\cite{Rin81} both for the initialization of the two partners of the reaction but also during the whole evolution.  The VAP technique itself is already the state of the art theory in nuclear structure and requires a large amount of computational resources. Beside, a proper formalism for its time-dependent equivalent (Time-Dependent VAP) where the many-body wave-function would be written as a set of evolving quasi-particle many-body vacua is still missing. A first attempt has been made in Refs \cite{Sca17,Sca17b} where the mixing is made approximately by assuming that each quasi-particle vacua evolves independently from each others. 
This attempt was the original motivation of the present work. We made extensive tests of this technique and realized that changing the 
phase convention during the TDHFB evolution significantly affects the results and therefore jeopardizes the reliability of the prediction.

\subsection{Semi-classical phase-space average over initial orientations}
\label{sec:semi}

A simpler approach toward the symmetry restoration relies on semi-classical averages over the initial gauge angle configurations. The attractive feature of this method is that it keeps the computational costs to the level of computing several mean-field trajectories. 

\subsubsection{Phase-space estimation of the moments of an observable}

The method starts with the statement that no orientation in gauge space should be \textit{a priori} privileged in the initial state. The situation is similar to the case of deformed nuclei where semi-classical methods with random orientations of deformation axis between nuclei have been considered followed by a set of classical evolutions to describe barrier fluctuations (see for instance~\cite{Esb78,Ayi10}).

Since the initial relative gauge angle $\theta^0_{AB}$ is arbitrary chosen before the two systems interact, one should at least perform a phase-space average all the possible orientations between $0$ and $2 \pi$. This is equivalent to assume a uniform initial probability for the relative gauge angle distribution:
\begin{eqnarray}
P(\theta^0_{AB}) &=& \frac{1}{2 \pi}. 
\end{eqnarray}
A simple observable $\hat O$ \footnote{Here simple observable means that its expectation value can be written as a linear combination of the one-body density matrix and anomalous density matrix elements.} is considered as a classical variable whose evolution is given by its expectation value along the mean-field path. These evolutions are denoted by ${\cal O}[\theta^0_{AB},t]$ since they explicitly depend on the initial relative orientation. 
In this picture, the quantum fluctuations in the gauge space are mimicked by the fact that ${\cal O}[\theta^0_{AB},t]$ becomes a random variable. The moments of order $k$ of the observable $\hat{O}$ after the collision is estimated through the semi-classical average:
\begin{eqnarray}
\overline{ O^k} & \equiv & \int_0^{2\pi} {\cal O}^k[\theta^0_{AB},t_\infty] P(\theta^0_{AB}) d  \theta^0_{AB},
\end{eqnarray}  
while its associated centered moment $\mu^{sc}_k$ reads:
\begin{equation}
\label{eq:scmoment}
 \mu^{sc}_k = \int_0^{2\pi} 
 \left(
 {\cal O}[\theta^0_{AB},t_\infty] - \overline{O}  
 \right)^k  P(\theta^0_{AB}) d\theta^0_{AB}
\end{equation}

This {\it brute-force} semi-classical treatment, that was already discussed extensively in Ref. \cite{Bro91}, is very similar to the stochastic mean-field  (SMF) approach \cite{Ayi08,Lac14} especially to its superfluid
version \cite{Lac13}. It should however be noted that here the initial phase-space is taken to restore in a classical picture the broken symmetry 
while in Ref.  \cite{Lac13} the initial phase-space was chosen to simulate quantum fluctuations of a quasi-particle vacuum or more generally of correlated systems. 
Besides being simpler technically, such a direct phase-space average has some advantages compared to 
alternative formulations where the quantum expectation are kept together with the gauge-angle average as in Refs. 
\cite{Cas97,Kur13}. Indeed, assuming an isolated Fermi superfluid with a fixed number of particles, 
all 
centered moments $\mu_k^{sc}$ of the total particle number distribution will be $0$ in the semi-classical picture while spurious fluctuations will persist if a quantum average is performed.

\subsubsection{Transfer probabilities from the phase-space approach}

In this section, we apply the phase-space averaging to compute the moments of the observable $\hat{X}=\hat{N}_A - N_A^0$ (number of pairs transfered to the system A). For this model case, we can compare the semi-classical results with the exact quantum distribution for the same observable. Indeed, once the exact solution has been evolved in time according to section~\ref{sec:exasol}, the exact centered moments for $\hat{X}$ can be evaluated with
\begin{equation}
 \mu^{exa}_k = \sum_x (x - \langle \hat{X}(t_\infty) \rangle)^k \, P^{exa}_{xn},
\end{equation}
where equation (\ref{eq:pexa}) provides the probabilities $P^{exa}_{xn}$ and sum runs over the even integers $x$.

The figure~\ref{fig:mom24a} compares the evolution of the second, fourth and sixth centered moments of $\hat{X}$ for a wide range of coupling strength. Both moments obtained from the exact case and the semi-classical phase-space approach are represented. 
In the strong coupling regime, significant differences between exact and semi-classical estimations are present for all three moments. 
In the low coupling regime, the situation is quite different. An impressive result is that the approximate and exact second moments coincide over a wide range of coupling strength. We see in particular (bottom part of Fig.~\ref{fig:mom24a}) that the ratio of the two moments is nearly constant and close to one up to $v_0/g \simeq 3.10^{-2}$ in the perturbative regime. In this weak coupling regime, the different centered moments of the number of nucleons in the sub-system $A$ can be expressed analytically by performing the averages of $N^k_A[\theta^0_{AB},t_\infty]$ over $\theta^0_{AB}$, where $N_A[\theta^0_{AB},t]$ is given by Eq.~(\ref{eq:nasym}). For the second moments, it gives for instance 
\begin{align}
\mu^{\rm sc}_2  & =  \overline{N^2_A} -   \overline{N_A}^2 =  \frac{1}{2} S^2_v (t_\infty) . \label{eq:muna2}
\end{align}
The fact that exact and semi-classical second order moments become proportional in the low coupling regime can in this case be shown analytically as:
\begin{align}
 \mu_2^{\rm sc} = \frac{1}{2}S_v^2(t_\infty) \propto \left(\frac{ v_0\tau_c}{\hbar}\right)^2 , ~\mu_2^{\rm ex} \simeq 8 P_{2n} \propto \left(\frac{v_0\tau_c}{\hbar}\right)^2.  \nonumber \\
\end{align}
The exact estimation of $P_{2n}$ comes from keeping only the leading order term in the time-dependent perturbative approach developed in the appendix~\ref{ap:exa}.

On the contrary,  higher moments obtained with the semi-classical average completely fail to reproduce their exact counterparts for all coupling strengths. In particular, the fourth and sixth semi-classical moments fall down much faster than the exact case when $v_0$ decreases. Using the analytical formula (\ref{eq:sv}), we can for example investigate the asymptotic behavior of the fourth semi-classical moment:
\begin{align}
 \mu^{\rm sc}_4 = \frac{3}{8} S_v^4(t_\infty) 
 \propto \left( \frac{v_0 \tau_c}{\hbar} \right)^4.
\end{align}
Due to the hierarchy of the probabilities $P_{2kn} \gg P_{2(k+1)n}$ in the perturbative region, the exact solution will necessarily result in a different behavior $\mu^{\rm ex}_4 \simeq 4 \mu^{\rm ex}_2 \propto  \left( \frac{v_0 \tau_c}{\hbar} \right)^2$. This relation between the order two and fourth exact moments clearly appears in the figure~\ref{fig:mom24a} and similar arguments can be used to explain the mismatch between exact and semi-classical moments for the higher orders.

\begin{figure}[h!]
 \includegraphics[width=0.95\linewidth]{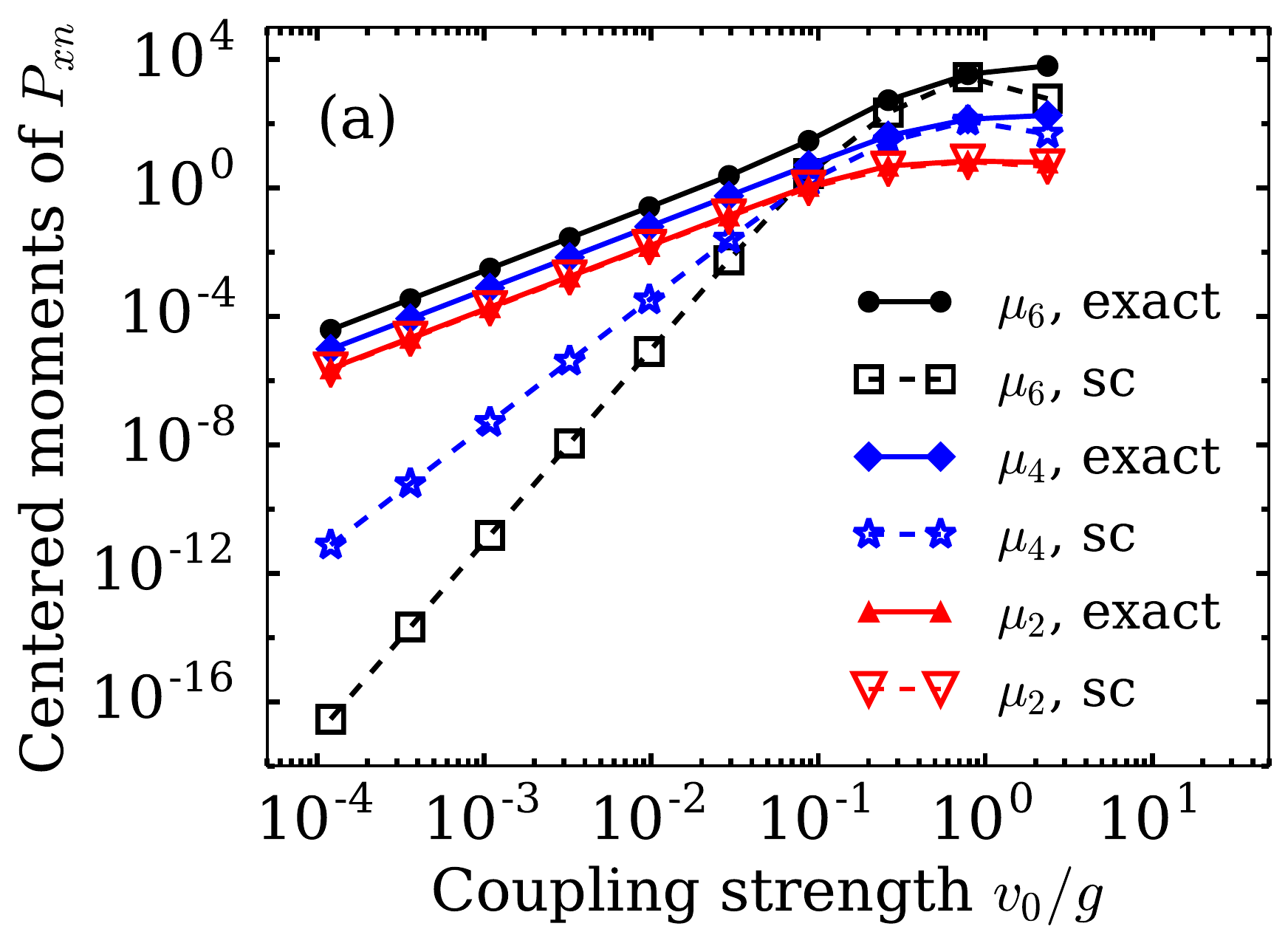}
 \includegraphics[width=0.9\linewidth]{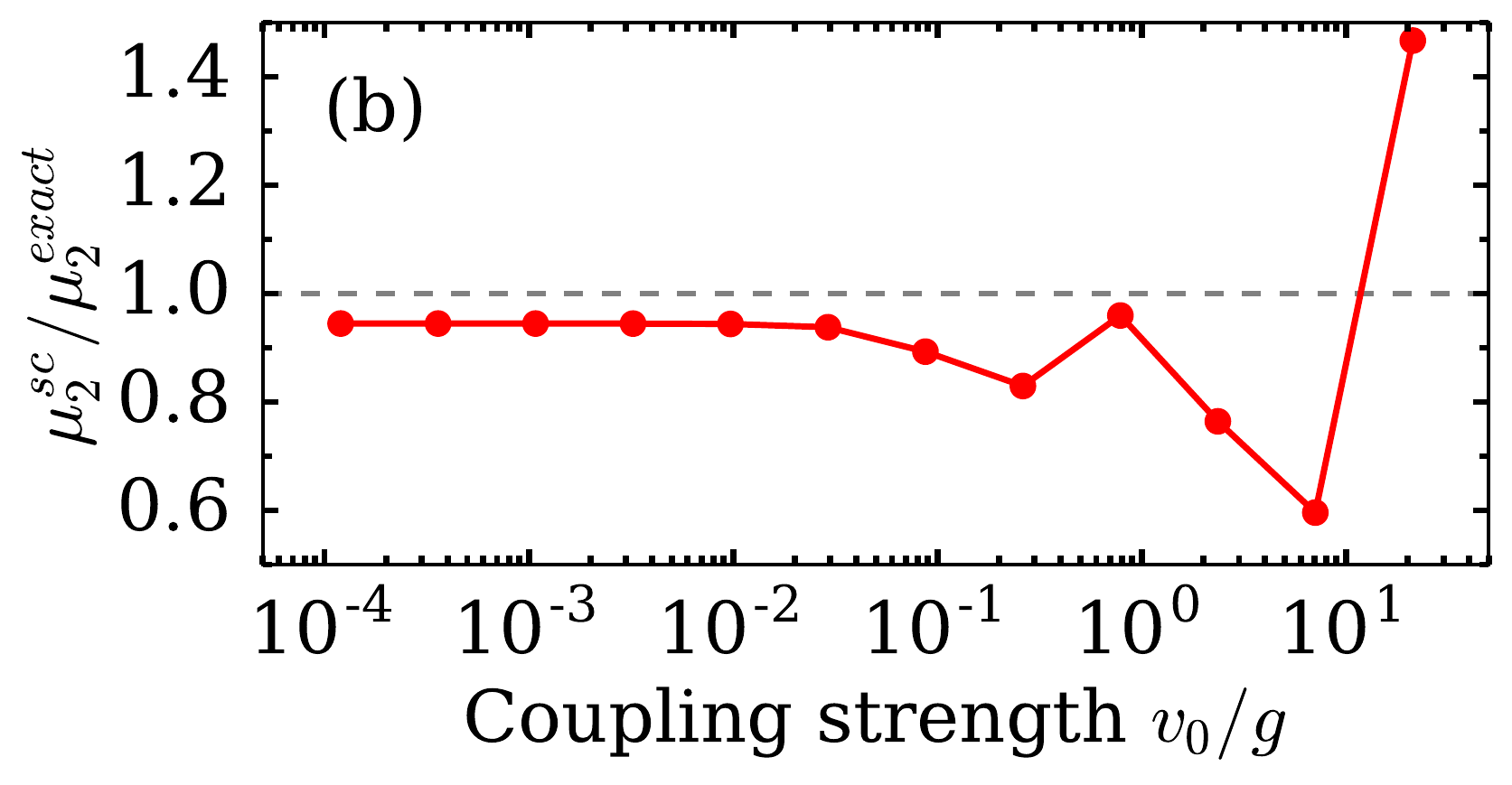}
 \caption{(a) Centered moments of the probability distribution $P_{xn}$ at final time as a function of  the coupling strength $v_0/g$. The semi-classical results (dashed lines)  are compared to the exact results (solid lines). (b) Second centered moment obtained with the semi-classical average divided by its exact counterpart. 
 The results are obtained for the symmetric degenerate case with $\Omega_A=\Omega_B=6$ and $N_A^0=N_B^0=6$. All TDHFB calculations are performed with a rescaling factor $\alpha=1.2$ on the interaction.}
 \label{fig:mom24a}
\end{figure}
%
As illustrated from the mismatch between moments of order higher than two, the distribution of the random variable $X[\theta^{0}_{AB}, t_\infty]$ is a poor approximation of the exact quantum distribution of the observable $\hat{X}$ at $t_\infty$. There are several reasons for the failure of a direct semi-classical phase-space average. First, semi-classical variable $X[\theta^{0}_{AB}, t_\infty]$ takes continuous real values whereas only even integers are possible in the exact treatment. In addition the domain of variation of $X[\theta^{0}_{AB}, t_\infty]$ may in practice be very different from the range of possible measurements of $\hat{X}$. To illustrate this, we emphasize that in the weak coupling regime one can eventually obtain an analytical expression of the probability $P^{sc}_{xn}$ probability using~\cite{Bro91}
\begin{equation}
 P^{sc}_{xn} =  \frac{1}{\pi} \left( \frac{d X[\theta^0_{AB}, t_\infty]}{d \theta^0_{AB}} \right)^{-1}.
\end{equation}
This expression
\footnote{
Note that this expression is only defined when $\frac{dX}{d\theta^0_{AB}} \neq 0$. According to equation this is not the case for a few discrete values of $\theta_0^{AB}$ (\textit{e.g} $\theta^0_{AB}=\pi/2$). This is actually not a problem as continuous distribution of probabilities may have non defined values on a support of measure null.}
accounts for the fact that two initial relative orientations lead to the same final number of particles (see Fig. \ref{fig:phase1}). 
Inserting the equation (\ref{eq:nasym}), we find that the number of particles is bounded
\begin{eqnarray}
N^0_A - S_v (t_\infty)  \le N^\infty_A \le N^0_A + S_v (t_\infty),
\end{eqnarray} 
which corresponds to the "classically allowed" region. In the exact treatment, particle number that are outside the classical region are always populated.

We conclude from this analysis that, even in our simple model case, the direct phase-space approach that consists in averaging over different TDHFB does not provide a precise description of the transfer probability distribution. However, this semi-classical average procedure seems to recover correctly the second moments of the pair transfer distribution by somehow including effects beyond the independent quasi-particle picture.

\subsubsection{Precision on the comparison between exact and phase-space treatment}
\label{sec:crit}
When comparing an approximate treatment of the many-body problem with the exact one, we should a priori use the same Hamiltonian. In Ref.~\cite{Sca17}, it has been argued that the interactions strength used in the Hamiltonian for the mean-field-based calculations should be rescaled compared to the exact case, in such a way that the initial total energy is the same. More precisely, for the symmetric case, it was assumed in \cite{Sca17} that $g'_A= \alpha g_A$,  $g'_B= \alpha g_B$ and $v_0'= \alpha v_0$ in the energy where $\alpha$ is a factor that depends on the specific case under study. 
For $\Omega_A=\Omega_B=6$ in the half filling situation we get $\alpha=1.2$ and the corresponding scaling was applied in the phase-space calculations presented here. This adjustment might appear surprising but it is quite close to what is done nowadays in nuclear structure where the strength of the pairing interaction is adjusted to reproduce either the pairing gap for mid-shell nuclei or the global trend in the two nucleons separation energies. It should however be kept in mind that this adjustment is based on rather empirical arguments. The comparison between the exact solution and approximate methods using this rescaling should be taken with caution and conclusions should only be qualitative.

\subsection{The phase-space combinatorial technique (PSC)}
\label{sec:transferproba}

Although the semi-classical technique presented in section~\ref{sec:semi} is not able to describe the complete richness of the transfer probabilities, it provides a very good estimation of the second moment for a wide range of interaction strength $v_0$.
This non trivial feature of the semi-classical approach for transfer was already emphasized in Ref.~\cite{Sca17} and is confirmed on a more systematic basis in this work.
Based on this empirical assessment, we propose a method to compute the pair transfer probabilities from the first semi-classical moments in the perturbative regime.
The method relies on two major assumptions:
\begin{itemize}
  \item[(a)] The first and second moments obtained from a semi-classical distribution of many TDHFB evolutions with different relative gauge-angles are realistic. This hypothesis can only be validated \textit{a posteriori} by comparing the result of the approach developed here with experimental observations.
  \item[(b)] The transfer of interest takes place in the weak coupling regime. In this regime, we do expect a hierarchy in the transfer probabilities.
\begin{eqnarray}
P_{0n} \gg (P_{2n}, P_{-2n}) \gg (P_{4n}, P_{-4n})  \gg \cdots  
\label{eq:hierarchy}
\end{eqnarray}
Such hierarchy is typically observed in reactions below the Coulomb barrier~\cite{Cor11,Mon14,Mon16,Raf16}, which establishes an interesting range of applications.
\end{itemize}


\subsubsection{One pair transfer}
\label{sec:onepair}

The one pair transfer is fully determined by the two probabilities $P_{2n}$ and $P_{-2n}$. 
For the symmetric case, we have in addition $P_{xn} \simeq P_{-xn}$. We can then use the hypothesis (b) to obtain an approximate expression between the variance of the distribution $P_{xn}$ and the two-particles addition/removal probabilities:
\begin{eqnarray}
\mu_2(t) & \simeq & 8 P_{2n}(t) = 8 P_{-2n}(t) . 
\label{eq:mu2p2}  
\end{eqnarray}  
We have checked that this is indeed realized up to 0.4\% in the exact calculations as long as we stay in the 
perturbative regime $v_0/g < 2.10^{-2}$. 
Consistently with the assumption made above, $P_{0n}$ is automatically obtained from the approximate relation:
\begin{eqnarray}
P_{0n} & \simeq& 1 - P_{2n} -P_{-2n}. \nonumber
\end{eqnarray} 
The relation (\ref{eq:mu2p2}) provides a straightforward way to access the two-particles addition/removal probabilities and avoids the complexity of multiple projections at different times \cite{Sca17}. In addition, it only requires the computation of independent mean-field trajectories. 
In the weak coupling regime between two symmetric degenerated systems, one can eventually use 
expression (\ref{eq:muna2}) to get the analytical form $P^{\rm sc}_{2n} = S^2_v(t)/16$ where $S^2_v(t)$ is given by Eq. (\ref{eq:sv}). 

\begin{figure}[ht!]
 \includegraphics[width=0.95\linewidth]{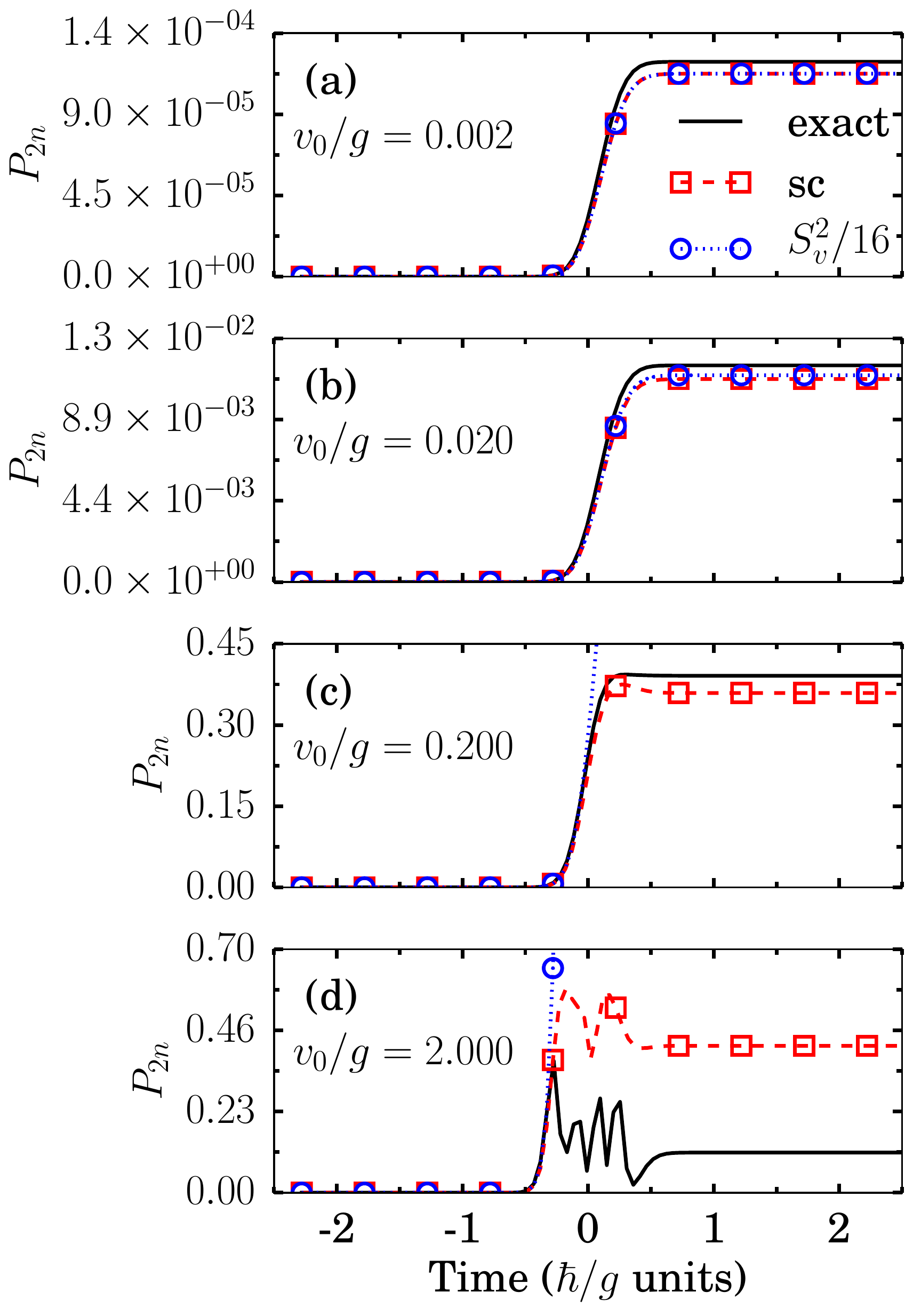}
 \caption{Two-particles transfer probability as a function of time and for different coupling strength $v_0/g$. 
 The semi-classical estimation (red dashed line) and its approximate analytical expression in the weak coupling regime (blue dotted line) are plotted along with the exact solution (black solid line). 
 The calculations are performed in the symmetric degenerate case with $\Omega_A=\Omega_B=6$ and $N_A^0=N_B^0=6$. The rescaling factor on the interaction for TDHFB calculations is $\alpha=1.2$.}
 \label{fig:p2time}
\end{figure}
In figure~\ref{fig:p2time}, the numerical and analytical semi-classical estimates of the two-particle pair transfer 
probabilities are compared to the exact ones as a function of time for different 
coupling strengths. At very small coupling, all the probabilities are in close agreement with each other. The simple strategy proposed here reproduces to a good extent the behavior of the exact result in the weak coupling regime. 
In particular, the time evolution of $P^{\rm sc}_{2n}(t)$ is smooth and  does not suffer from the spurious oscillations of asymptotic probabilities observed in Refs.~\cite{Sca17,Sca17b}. As could be anticipated, the numerical estimate has a wider range of applicability than the analytical one obtained from $S^2_v(t)/16$. 
Finally, and without surprise, more and more deviation is observed as the coupling strength enters the non-perturbative regime.   

\subsubsection{From one to multiple pair transfer}

The success of the above method is an incentive to generalize it to multiple pair transfer. A naive attempt in this direction would be (i) to compute the higher order semi-classical moments $\mu^{sc}_k$ (ii) to invert the set of equations $\mu^{sc}_k =  \sum_{n} (2n)^k (P_{2kn}+ P_{-2kn})$ to retrieve the probabilities. This technique would actually work if the high order moments $\mu^{sc}_k$ would match their exact counterparts. However, the figure~\ref{fig:mom24a} clearly shows that the phase space approach fails to predict the centered moments of order higher than $2$.
 
To circumvent this difficulty, we propose to model the shape of the probability distribution $P_{xn}$ with an analytical formula involving a sufficiently small number of free parameters. Once the generic shape of the distribution is decided, we determine the parameters so to reproduce the first and second moments predicted by the semi-classical average. This gives us eventually the possibility to extrapolate the distribution $P_{xn}$ to multiple pair transfer.

To propose a shape for the $P_{xn}$ distribution, we start from the following simplifying assumption.
\begin{itemize}
  \item[(c)] In the perturbative regime,  the transfer of several pairs from one superfluid system to another 
  can be essentially treated as a sequence of uncorrelated pair transfers.   
   \end{itemize}
This hypothesis is guided by the exact resolution that is discussed  in appendix~\ref{ap:exa}. In the exact case, 
the transfer probabilities result from a rather complicated process involving interferences between different 
channels (see Fig.~\ref{fig:interference} of appendix~\ref{ap:exa}).  In the perturbative regime, the interference between 
channels can be neglected and we simply end-up with a sequence of transfer that could be depicted for the  addition
or removal  process from $A$ to $B$ respectively as:
\begin{eqnarray}
N_A \xrightarrow{{\cal P}_{+1}} N_A+2 \xrightarrow{{\cal P}_{+2}} N_A+4 \cdots  \label{eq:schem+}
\end{eqnarray} 
and
\begin{eqnarray}
N_A \xrightarrow{{\cal P}_{-1}} N_A-2 \xrightarrow{{\cal P}_{-2}} N_A-4 \cdots 
\end{eqnarray} 
The removal and addition probabilities of $k$ pairs can then be written as a product:
\begin{eqnarray}
P_{2kn}   & = & {\cal P}_{+1} \cdots {\cal P}_{+k}, ~~ P_{-2kn}    =  {\cal P}_{-1} \cdots {\cal P}_{-k}.
\end{eqnarray}
A simplified expression of the ${\cal P}_{+k}$ is illustrated by Eq. (\ref{eq:pksimp}). These transfer 
probabilities are in general rather complex since they contain the information on the internal structure of the system before and
after the transfer, as well as the dynamical effects of the time-dependent interaction. However, as shown in appendix~\ref{ap:exa},
 in some limiting situation, the product of probabilities can be rewritten as:
 \begin{eqnarray}
P_{2kn} & = & W_k p^k, ~~
P_{-2kn}  =   W_{-k} q^k,
\label{eq:simplepk}
\end{eqnarray}
where $W_{k/-k}$ are combinatorial factors while $p$ and $q$ can be interpreted 
as the elementary probability for a pair to be added or removed during the reactions. In the exact 
case,  this probability is governed by the coupling $v_0$ as well as the contact time, \textit{i.e.} by the time-dependent interaction between the
two systems. The factors $ W_k $ and $W_{-k}$ contain here the information on the available number of particles to transfer as well as the number of possible states reachable when accounting for the Pauli principle. It may also contain in an approximate way some information on the intrinsic structure of the initial and final states. In appendix~\ref{ap:exa}, we obtained the expression of $W_{k/-k}$
for two specific cases:
\begin{itemize}
  \item We first consider the case where the transition frequencies at each step of the process (\ref{eq:schem+}) are constant. Then, we have:
  \begin{eqnarray}
W_k  & = & C^k_{\Omega_A - n_A} C^k_{n_B}   \frac{(n_A + k)! }{n_A!}  \frac{ (\Omega_B - n_B + k )! }{(\Omega_B - n_B)!} .  \label{eq:wkcst}
\end{eqnarray} 
This situation is the one anticipated for the case of degenerate system {\it in the absence of pairing}, i.e. when both systems are in their normal phase.     
  \item We then consider the case where the transition frequencies at the step $k$ is proportional to $k$, leading to:
  \begin{align}
 W_{k}   =  C^k_{\Omega_A - n_A} C^k_{n_B} C^k_{n_A + k} C^k_{\Omega_B -n_B +k}.  
 \label{eq:wklin}
\end{align}
This second case, is relevant for two degenerate systems {\it in the presence of superfluidity} starting from half-filling 
and is then expected to be more realistic for the present study.\end{itemize}  
In both cases, $W_{-k}$ is deduced from $W_k$ simply by making the replacement $\Omega_A \leftrightarrow \Omega_B$ and 
$n_A  \leftrightarrow  n_B$.    Parts of the combinatorial factors appearing in Eqs.~(\ref{eq:wkcst}) and (\ref{eq:wklin}) have a simple statistical interpretation. Indeed, $C^k_{n_B}$ counts the number of configurations of $k$ pairs initially in system $B$ that could be transferred to $A$
while $C^k_{\Omega_A - n_A}$ counts the number of possibilities to put $k$ pairs in $\Omega_A - n_A$ empty spots. 
Note finally that, this second estimate of the pair transfer 
probabilities will lead to lower probabilities since Eq. (\ref{eq:wklin}) can be obtained by dividing Eq. (\ref{eq:wkcst}) by $(k!)^2$.  
The approach that uses combinatorial arguments is called hereafter PSC (Phase-Space Combinatorial). 

In the following, we will systematically show the result of both prescriptions.   Since both are deduced from simple approximate treatment of 
the internal structure of the systems, the variation of transfer probabilities from one prescription to another gives us an idea on the uncertainty 
related to the proper treatment of the structure of the two systems.


Our starting point to obtain multiple pair transfer probabilities are the assumed formula $(\ref{eq:simplepk})$ where we see that the 
important quantities for pair transfer addition and removal are the elementary probabilities $p$ and $q$. These parameters can directly be inferred 
using our previous technique to estimate two-particle transfer probabilities from phase-space average (section \ref{sec:onepair}). Focusing first on the symmetric case where $p=q$, we obtain simply:
\begin{eqnarray}
p=q= \frac{\mu_2}{8W_1}.
\end{eqnarray}
From the knowledge of $p$ and $q$, we can directly calculate the different probabilities when more than one pair
is transferred using Eqs. (\ref{eq:simplepk}). It is worth mentioning that for the symmetric case, we have 
the recurrence relation:
\begin{eqnarray}
P_{2 (k+1)n} & =& p \frac{W_{k+1}}{W_{k}} P_{2kn}, 
\label{eq:recurence}
\end{eqnarray}
as well as a direct connection between the probability to transfer $k$ pairs with the probability to transfer one pair:
\begin{eqnarray}
 P_{2kn} & = &  \frac{W_k}{W^k_1} \left[ P_{2n}\right]^k, ~~P_{-2kn}  = \frac{W_{-k}}{W^k_{-1}} \left[ P_{-2n}\right]^k .  \label{eq:scal}
\end{eqnarray}

\begin{figure}
 \includegraphics[width=0.9\linewidth]{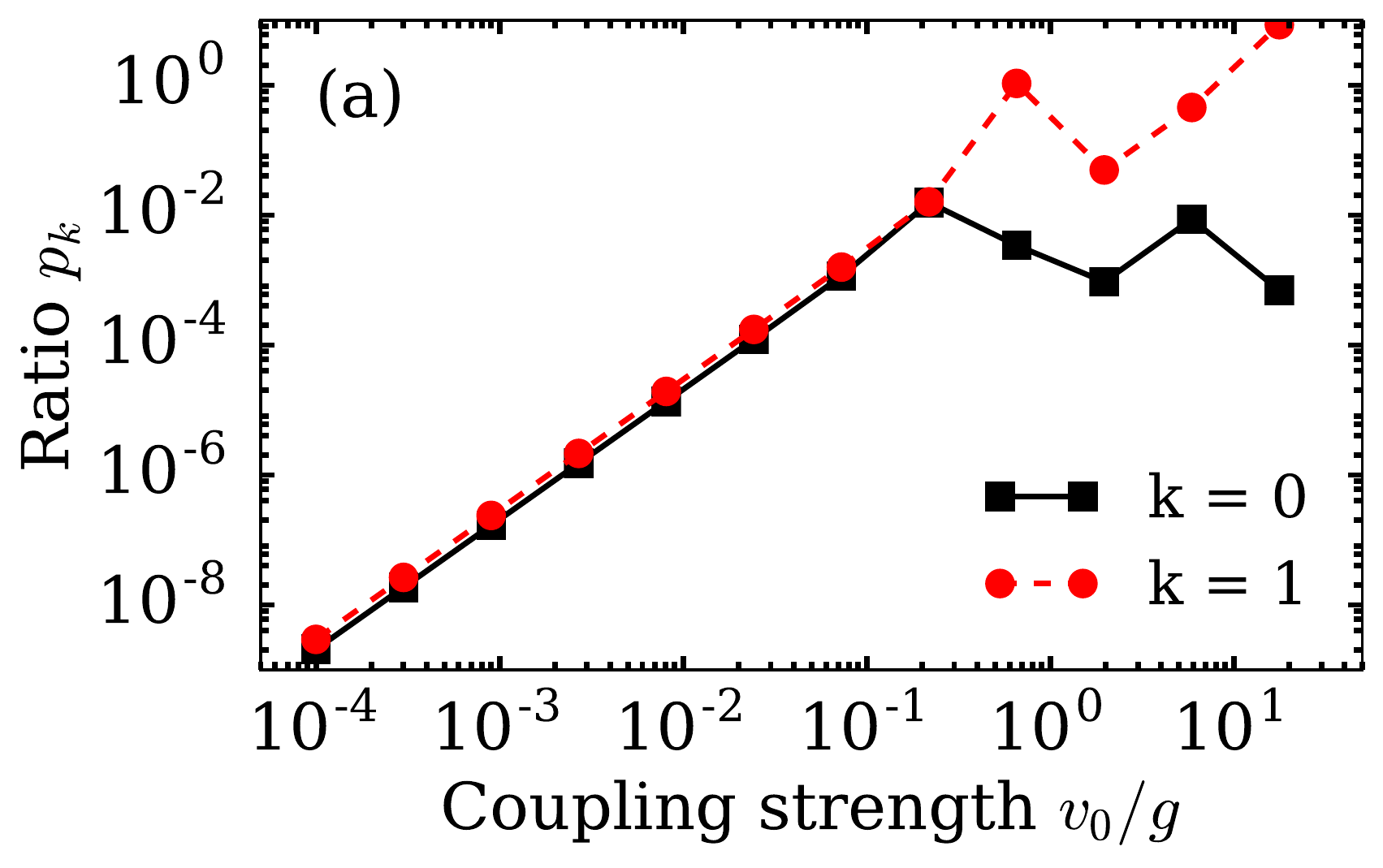} \\
 \includegraphics[width=0.9\linewidth]{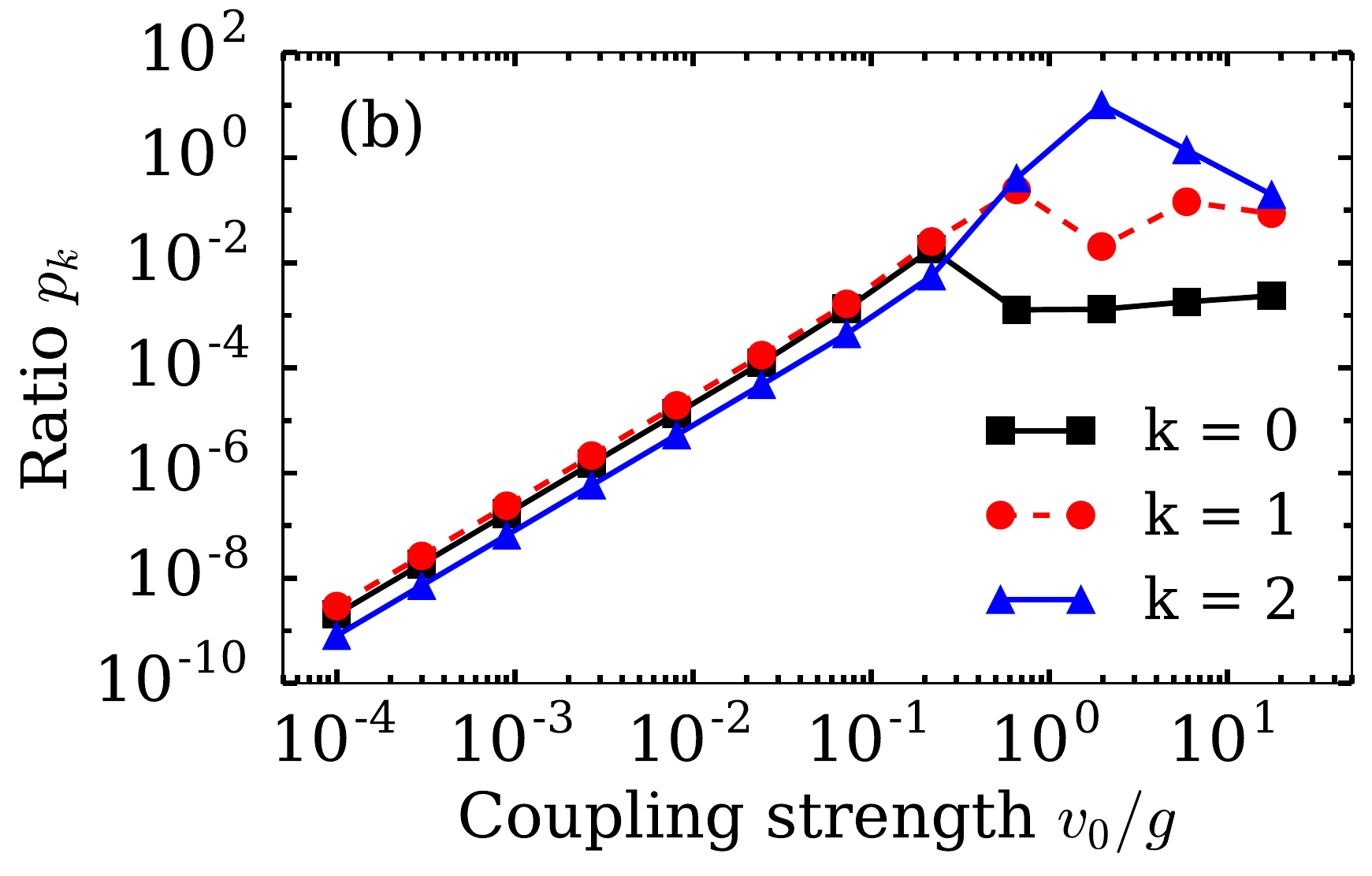}\\
 \includegraphics[width=0.9\linewidth]{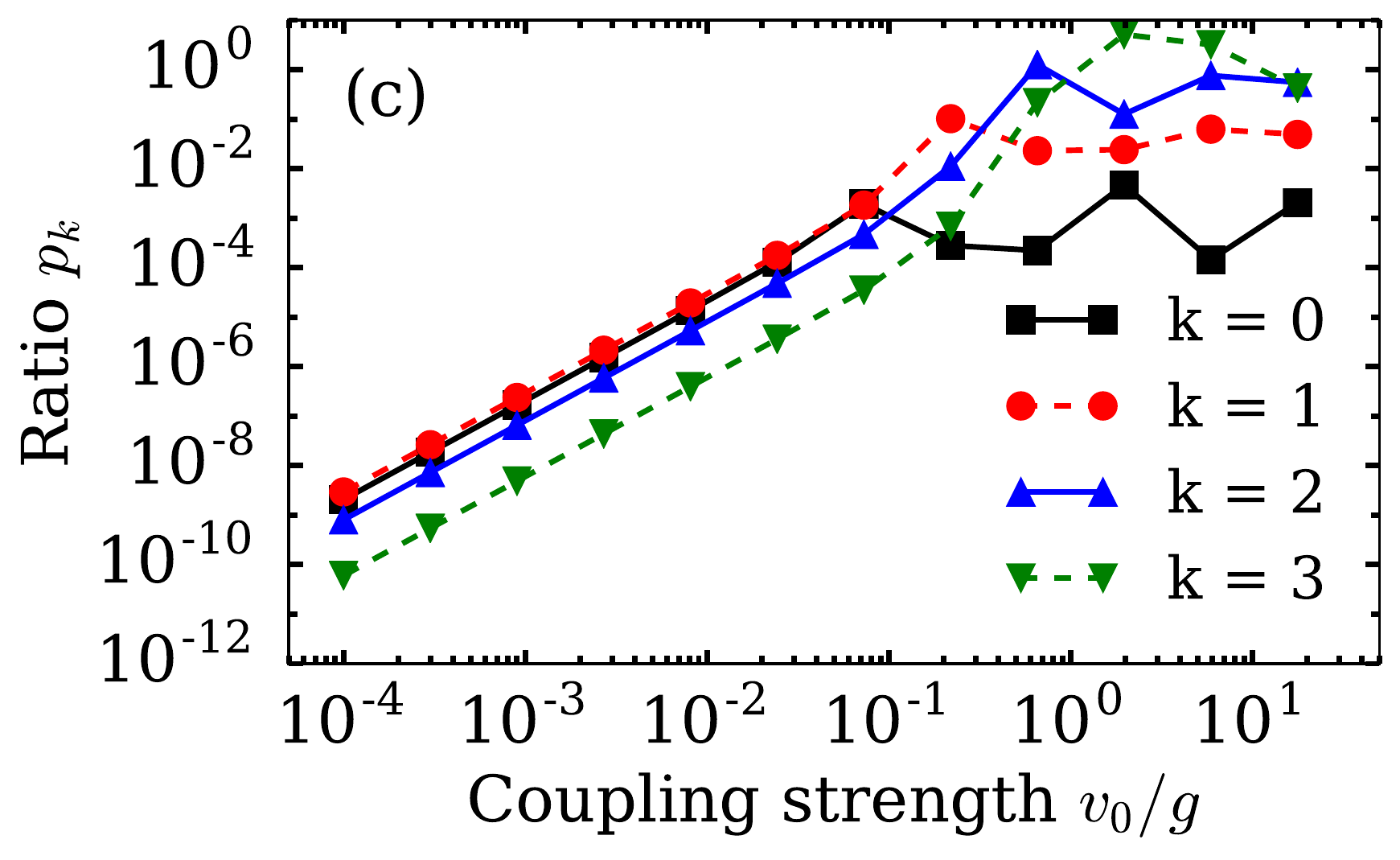}
 \caption{Ratios $p_k = [W_kP^{\rm ex}_{2(k+1)n}]/ [ W_{(k+1)} P^{\rm ex}_{2kn} ]$ 
 as a function of the coupling strength. 
 We used the $W_k$ coming from Eq.~(\ref{eq:wklin}). 
 The results are displayed for the symmetric degenerate case with (a) $\Omega_A=\Omega_B=4$, (b) $\Omega_A=\Omega_B=6$ and (c) $\Omega_A=\Omega_B=8$ starting from the half filling configuration for both systems.}
 \label{fig:p1mp}
\end{figure}
The method proposed here provides a straightforward way to obtain the transfer probabilities  from a semi-classical average using solely the second moment of the simulated distribution. Let us now study to what extent the forms given by Eqs.~(\ref{eq:simplepk}) are valid for the considered model. To do so, we compute the ratios  $p_k = [W_kP^{\rm ex}_{2(k+1)n}]/ [ W_{(k+1)} P^{\rm ex}_{2kn} ]$ obtained from the exact probabilities. The shape of the probability distribution is realistic when these ratios are almost independent of $k$ for a given value of  $v_0/g$ (cf. equation (\ref{eq:recurence})).
The figure~\ref{fig:p1mp} highlights these ratios for different $k$ and different initial size of the systems $A$ and $B$. We used the values of $W_k$ given by Eq. (\ref{eq:wklin}). All ratios match to a good extend with each others as long as not too many pairs are transferred. Note that, if the prescription (\ref{eq:wkcst}) is used, more deviations 
between ratios are observed as anticipated. 

\begin{figure}[h!]
 \includegraphics[width=0.9\linewidth]{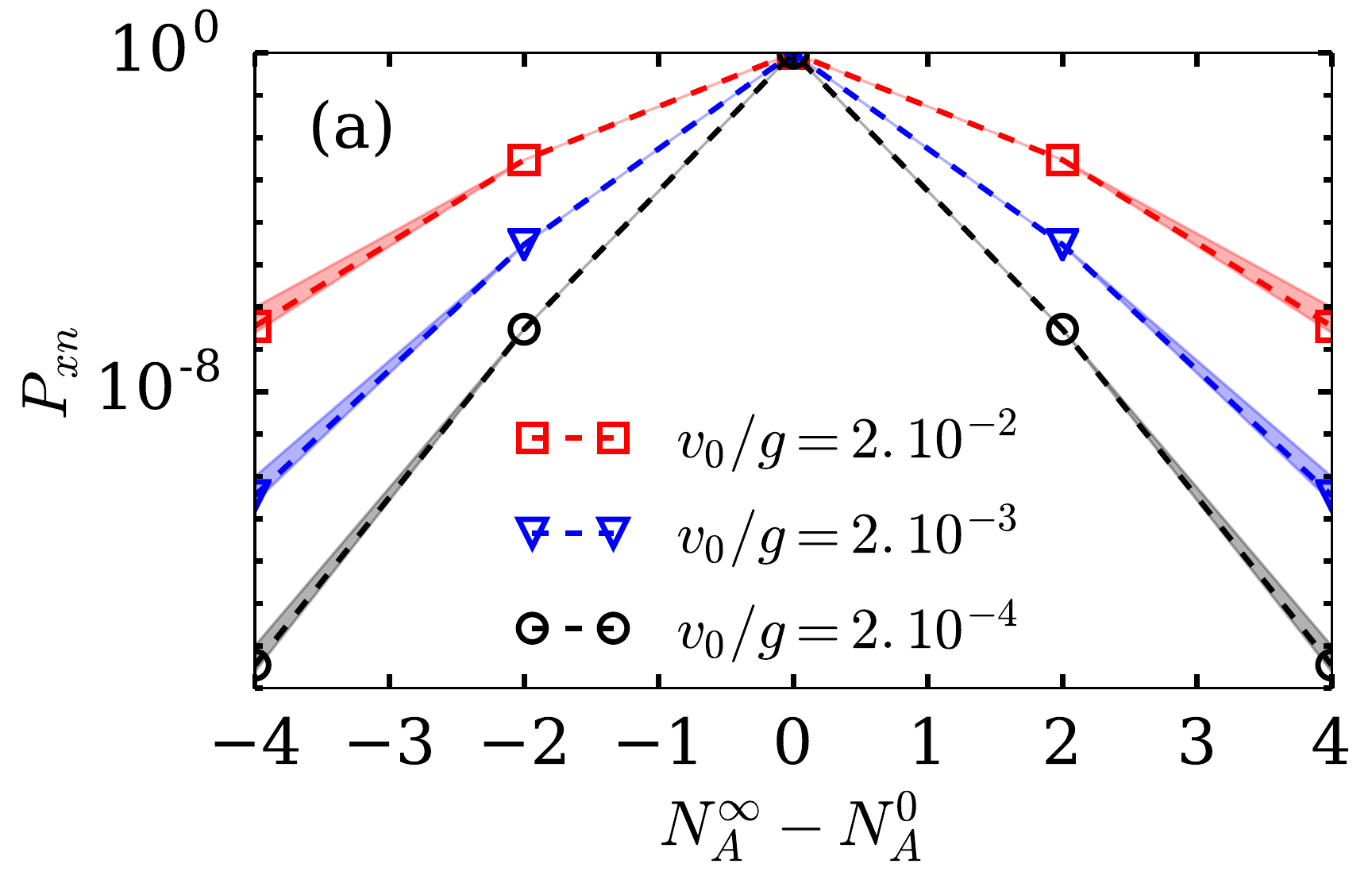}
 \includegraphics[width=0.9\linewidth]{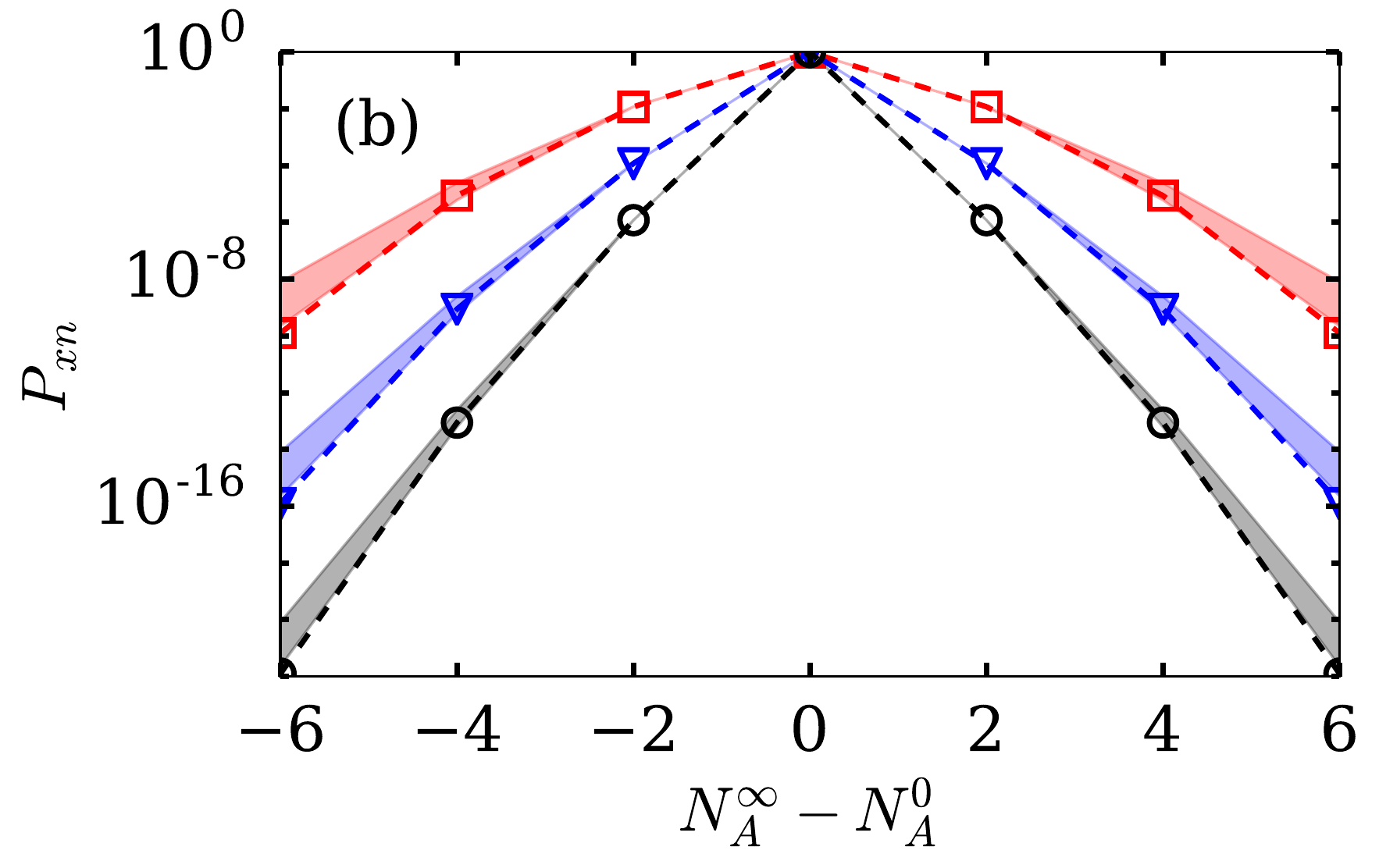}
 \includegraphics[width=0.9\linewidth]{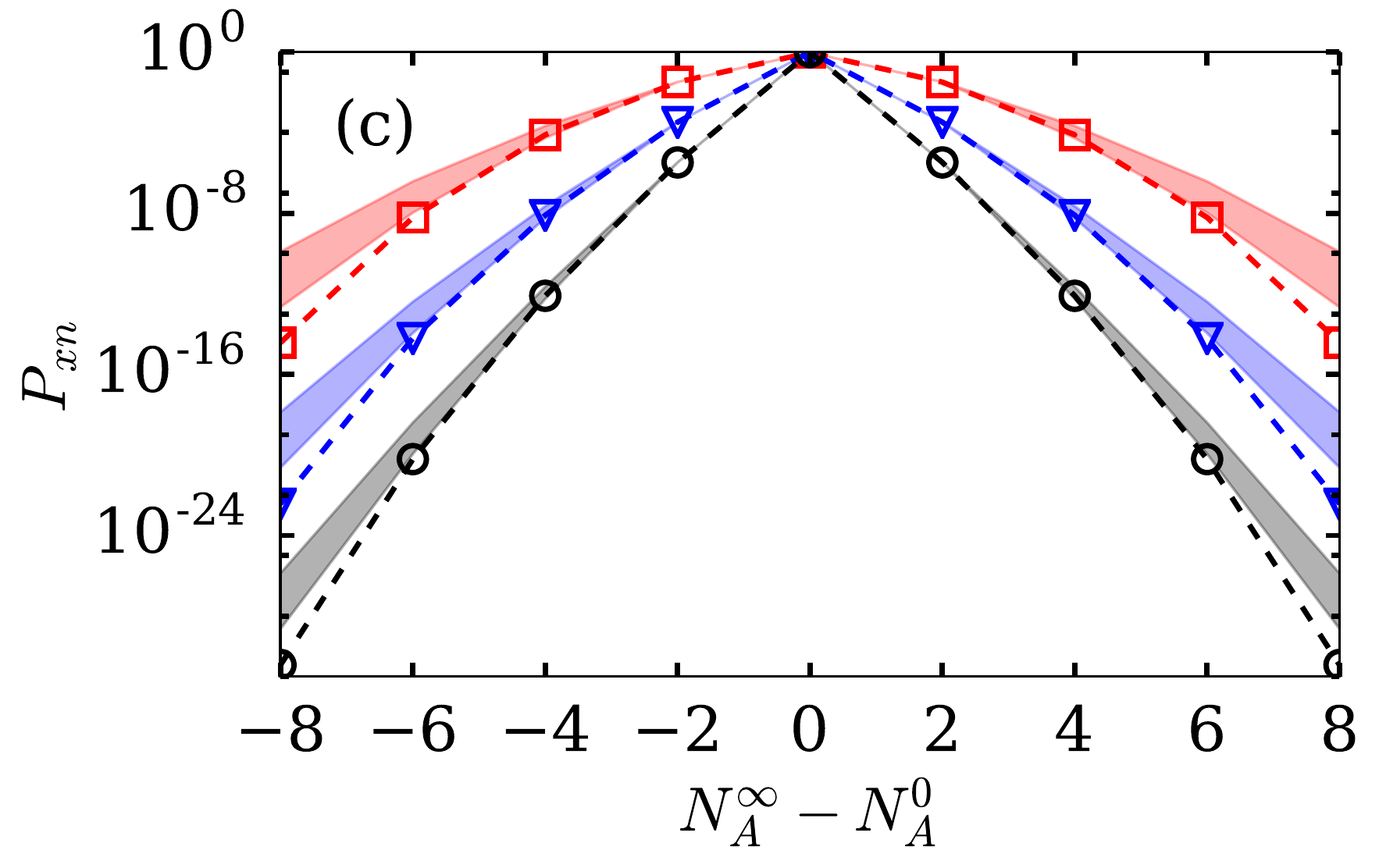}
 \caption{Asymptotic probabilities to transfer one or several pairs as a function of  the quantity 
 $N_A^{\infty}-N_A^{0}$ obtained from a symmetric degenerate situation with $\Omega_A=\Omega_B=N_A^0=N_B^0=$ 4 (a), 6 (b) and 8 (c). The colored bands span the area obtained with the distribution Eq.~(\ref{eq:simplepk}) with the two cases for the $W_k$ factors. The distribution are obtained by using the exact second moment $\mu^{\rm ex}_2$ 
 to determine $p=q$.  Approximate probabilities are systematically compared with the exact values (dashed line) for three coupling strength $v_0/g$ in the perturbative regime.}
 \label{fig:distrib_exact}
\end{figure}
To further illustrate the accuracy of the PSC approach, we show in Fig.~\ref{fig:distrib_exact}
the approximate pair transfer probabilities deduced when using the exact second moment $\mu^{\rm ex}_2$ in the combinatorial method presented here and compare it to the exact probabilities. 
We see that the pair transfer are relatively well reproduced with more and more deviations as the number of transferred pairs or as the coupling strength increases. Nevertheless, in general, the probability to transfer two pairs and sometimes three pairs are reasonably close to the exact probabilities.
\begin{figure}[h!]
 \includegraphics[width=0.9\linewidth]{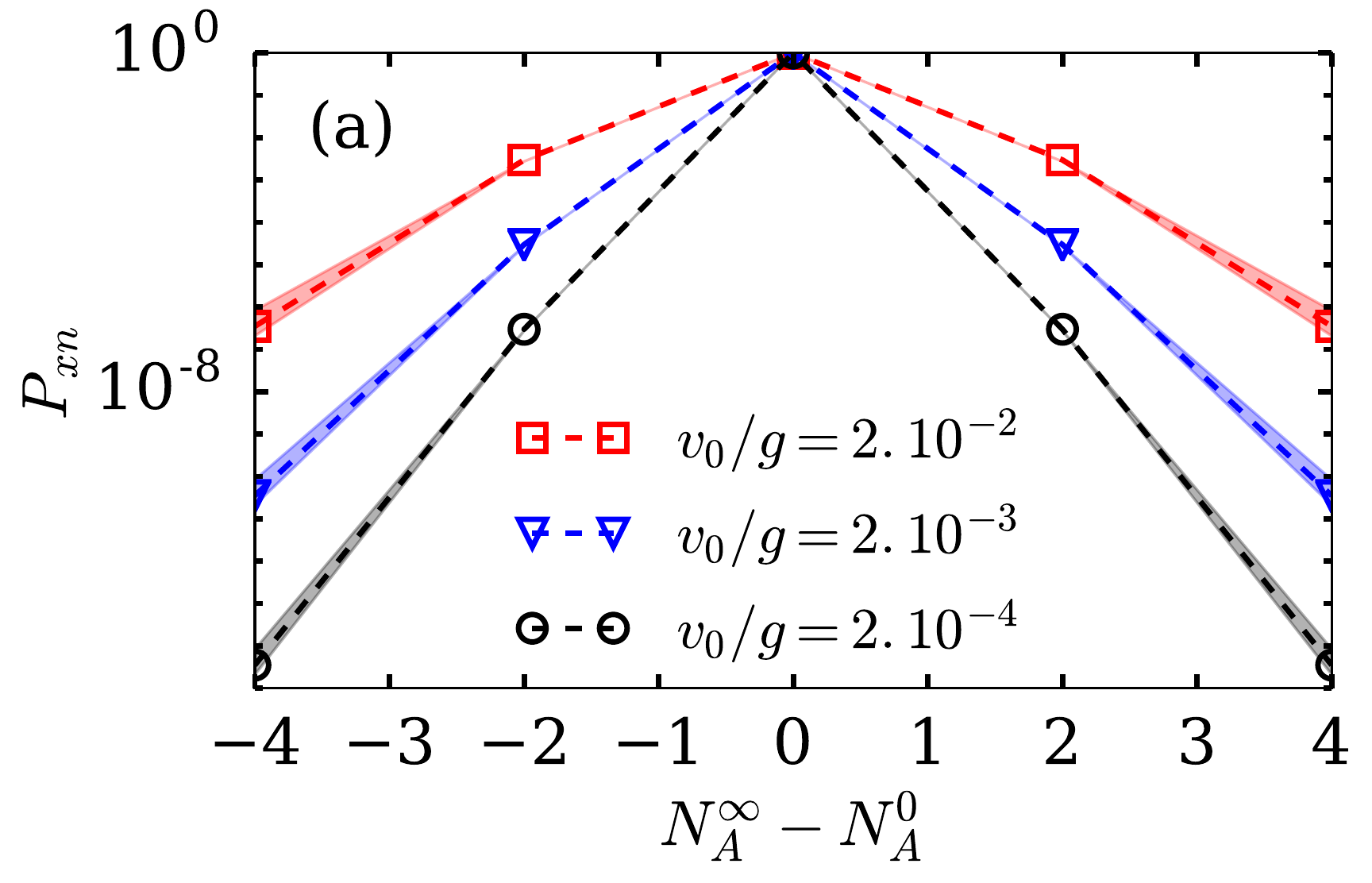}
 \includegraphics[width=0.9\linewidth]{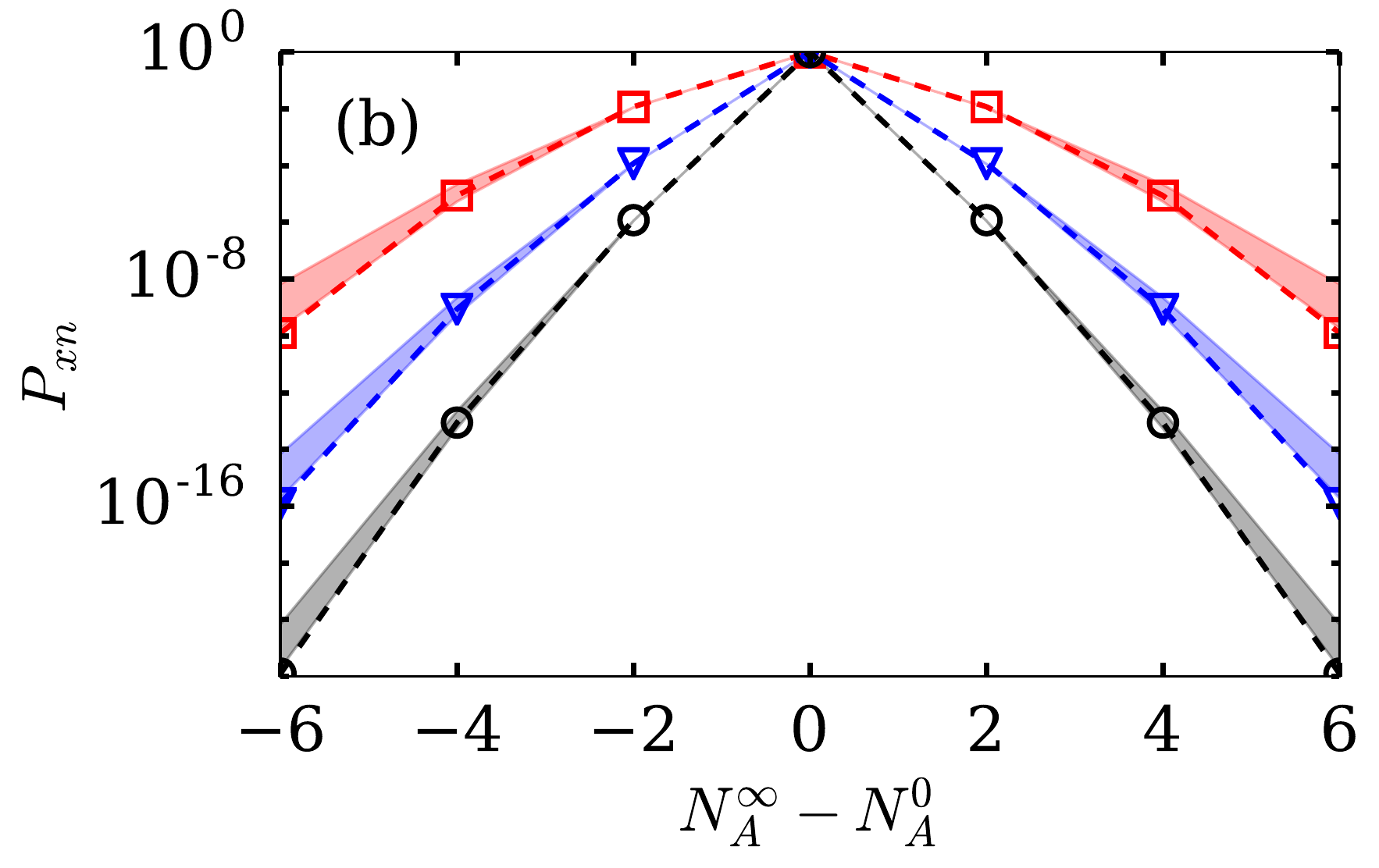}
 \includegraphics[width=0.9\linewidth]{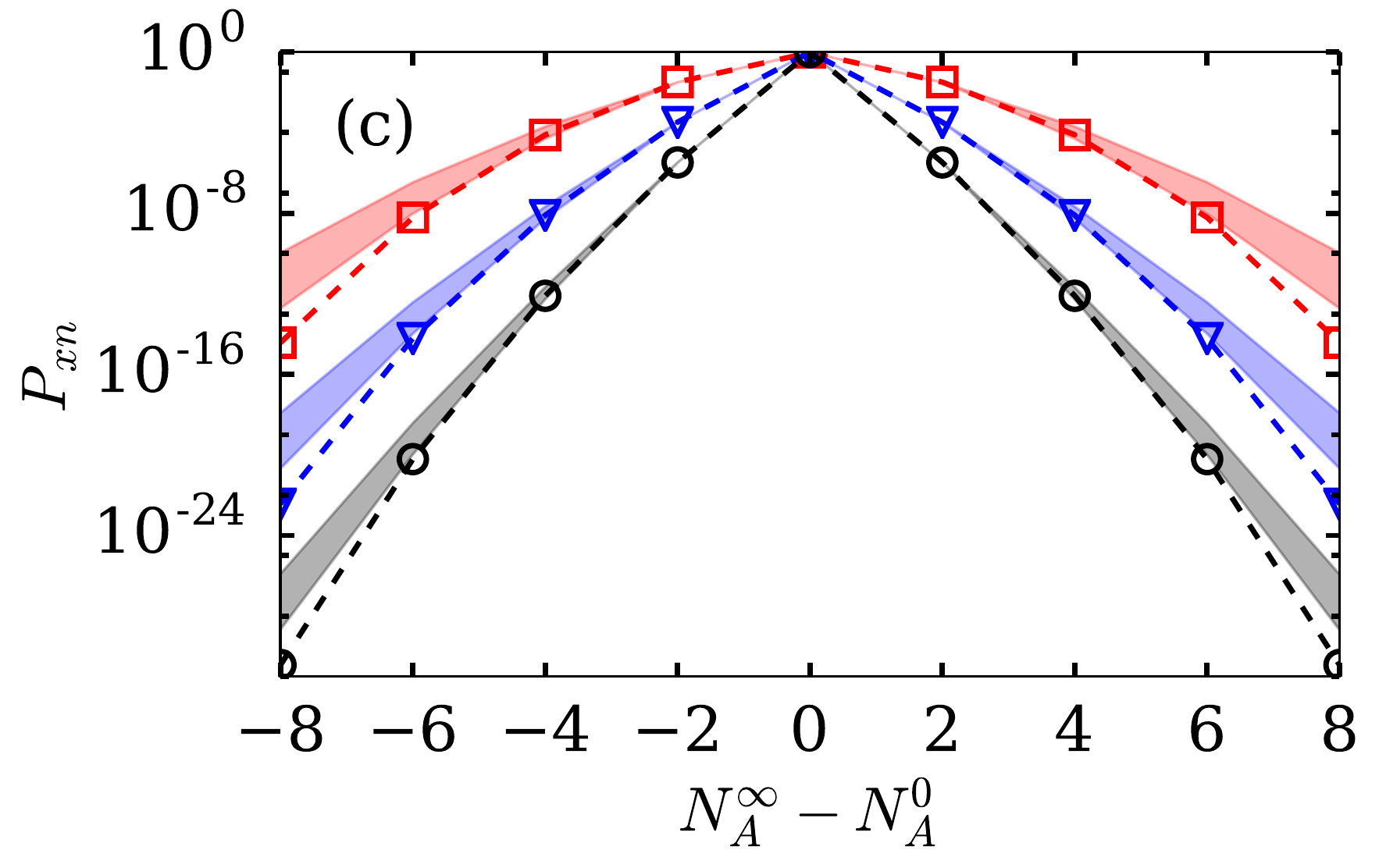}
 \caption{Same as figure \ref{fig:distrib_exact} except that $p$ is  determined with the second moment $\mu^{\rm sc}_2$ obtained  by performing the semi-classical phase-space average.}
 \label{fig:distrib}
\end{figure}
The figure~\ref{fig:pv0_tdbcs_sym} presents the systematic calculation of the multiple pair transfer as a function of the coupling strength $v_0/g$ in the perturbative regime. We see that the behavior of the distribution is monotone in this region and the semi-classical result gives a nice estimate of the exact probabilities in this case.
\begin{figure}[h!]
 \includegraphics[width=0.9\linewidth]{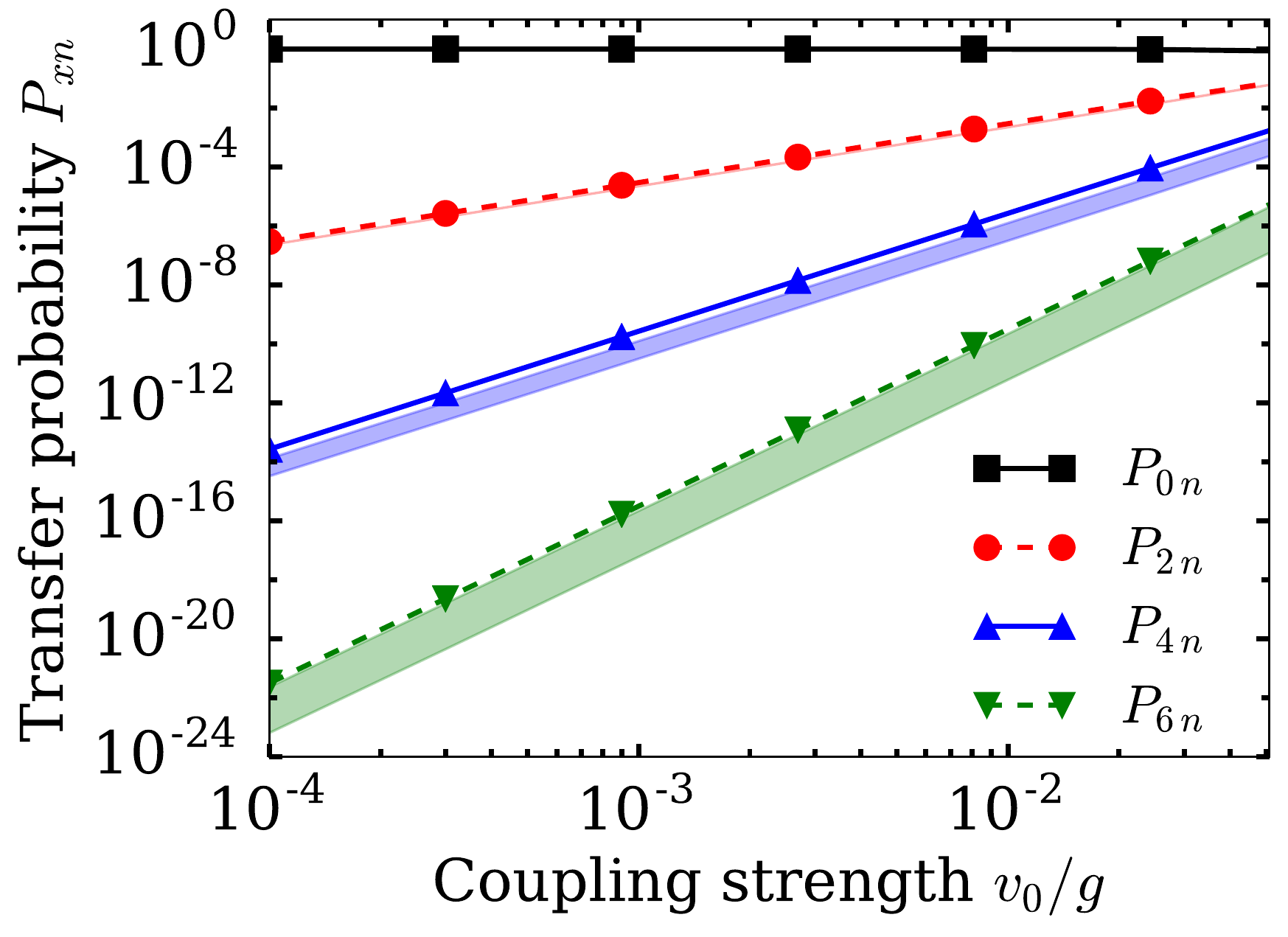}
 \caption{Asymptotic probabilities to transfer one or several pairs as a function of  the coupling strength $v_0/g$ and obtained from a symmetric degenerate situation with $\Omega_A=\Omega_B=N_A^0=N_B^0=$ 6. The colored bands span the area obtained with the distribution Eq.~(\ref{eq:simplepk}) with the two cases for the $W_k$ factors. The distribution are obtained by using the semi-classical second moment $\mu^{\rm sc}_2$ to determine $p=q$. Approximate probabilities are compared with the exact values (dashed line). All TDHFB calculations are performed with the rescaling factor $\alpha=1.2$.}
 \label{fig:pv0_tdbcs_sym}
\end{figure}

We finally couple the combinatorial approach with the semi-classical estimate of the second moment. Provided that the phase-space averaging is accurate to describe this moment (hypothesis (a)) we compute in a straightforward way the different pair transfer probabilities starting from $\mu^{\rm sc}_2$. The figures~\ref{fig:distrib} and ~\ref{fig:pv0_tdbcs_sym} compare again the approximate transfer probabilities using $\mu^{\rm sc}_2$ with the exact ones for the symmetric case. 
The overall evolution of the probability with increasing number of pairs transferred is rather well reproduced as well as its $v_0$ dependency. As expected, the exact solution is closer to the case of linear transition frequencies given by Eq.~(\ref{eq:wklin}). 
Note that in all cases, we have $\mu^{\rm sc}_2 \sim \mu^{\rm ex}_2$. Therefore, using the exact or semi-classical second moments  does not make so much differences in the estimated probabilities. 
Part of this matching is a direct consequence of the fact that the interaction has been rescaled in the TDHFB case to match the exact ground state energy (cf. \ref{sec:crit}). In the different cases considered  in Fig.~\ref{fig:distrib}, a scaling factor  $\alpha=1.333$, $\alpha=1.2$ and $\alpha=1.144$ has been used for $\Omega_A=$4, 6 and 8 respectively leading to a difference between the exact and approximate $\mu_2$ that is at maximum $10 \%$. Again, we would like to insist on the fact that the scaling procedure is irrelevant for realistic applications and one should instead suppose that the hypothesis (a) is valid.

In summary, we introduced here the new PSC method that works in two steps. First the second moments $\mu^{sc}_2$ is estimated from phase-space averaging. This requires computing a set of independent TDHFB trajectories. Then the complete pair transfer distribution is recovered from the combinatorial factors $W_k$. The main benefit is that we obtain beyond mean-field fluctuations at the cost of several mean-field calculations only. The growth of the computation time with the number of particles is therefore no more than the one associated to the TDHFB calculations themselves.


\section{Application to $^{20}$O+$^{20}$O reactions below the Coulomb barrier} 
\label{sec:o20o20}
 
To test the applicability the PSC method to a realistic situation, we consider the symmetric reaction $^{20}$O+$^{20}$O discussed in  Ref.~\cite{Sca17}. The TDHFB equation using the Gogny interaction \cite{Has12,Has13} has been used to simulate the central collision of two $^{20}$O superfluid nuclei 
at various energies and gauge angles below the Coulomb barrier. In~\cite{Sca17b}, a multiple projection technique was used to extract two particle transfer probabilities. 
With the same projection technique, the probability to transfer several pairs can also be obtained and can serve as an element 
of comparison to the PSC approach  proposed here. The one pair and two pairs transfer probabilities obtained
with projections are represented in Fig.~\ref{fig:o20o20} by solid lines. Note that in this case, the error bars 
stem from the fluctuations of the probabilities after the nucleus do re-separate.  

An important and non-trivial ingredient to be able to apply the PSC approach is to figure out the number of pairs contributing to the transfer and the size of the phase-space available for particles to be transmitted to the other nucleus. 
This number does not impact the probability of one pair exchange, but it drives the predictions for the multiple pairs transfer.
In a simple shell-model picture, for the $^{20}$O, we expect to have 4 particles in the last occupied level $1d_{5/2}$, therefore we assumed 
that the number of pairs equals $n_A=n_B= 2$. The available phase-space after transfer is more difficult to identify, 
a reasonable assumption is to suppose that all states in the $sd$ shell contribute to the phase-space, i.e. 
$(\Omega_A=\Omega_B=6)$.  We show in Fig.~\ref{fig:o20o20} the result of the combinatorial approach compared to the result obtained 
by projection. It is first remarkable to notice that the probability to transfer two particles are almost identical in the two approaches. This 
is already a great success of our approach in view of its relative simplicity compared to the method proposed in Ref.~\cite{Sca17b}. 
Indeed, the two-particle transfer is obtained in the present work using the simple formula (\ref{eq:mu2p2}) once the second moment 
is computed from the phase-space average.  Another advantage is the absence of dependence of the result with respect to the phase convention used to solve the TDHFB equations. Note that, the agreement of the two particles transfer 
indirectly validate the projector approach to estimate $P_{2n}$ and $P_{-2n}$.

The situation is different when more than one pair is transferred.  We see that the PSC approach leads systematically to probabilities that are smaller than the projection case. 
%
%
One possible origin of the discrepancy could stem from the size of the available phase-phase that is assumed to perform the PSC calculation. Indeed, the multiple pair transfer increases with degeneracy $\Omega_A$. To obtain an upper limit of $P_{4n}$ with the PSC approach, we simply assumed an infinite number of possible final states, \textit{i.e.} $\Omega_A=\Omega_B=+\infty$.
Using Eqs. (\ref{eq:wkcst}) and (\ref{eq:wklin}) lead to the scaling $P_{4n} = 1/6 \left[ P_{2n}\right]^2$ and  
$P_{4n} = 1/24 \left[ P_{2n}\right]^2$ respectively. These upper limits are actually within 10\% of the values obtained with $\Omega_A=6$ which tells us that (i) the PSC is quite robust relatively to a change of $\Omega_A$, (ii) having more states active in the reaction does not explain the discrepancy with the projection case where the scaling $P_{4n} \simeq 1/2 \left[ P_{2n}\right]^2$ has been empirically found. Having in mind the success of the approach in the toy model 
and henceforth supposing that the approach is suited to predict the multiple pair transfer, this discrepancy suggests that the projection technique overestimates the probabilities when more than one pair is transferred.

\begin{figure}[h!]
 \includegraphics[width=0.9\linewidth]{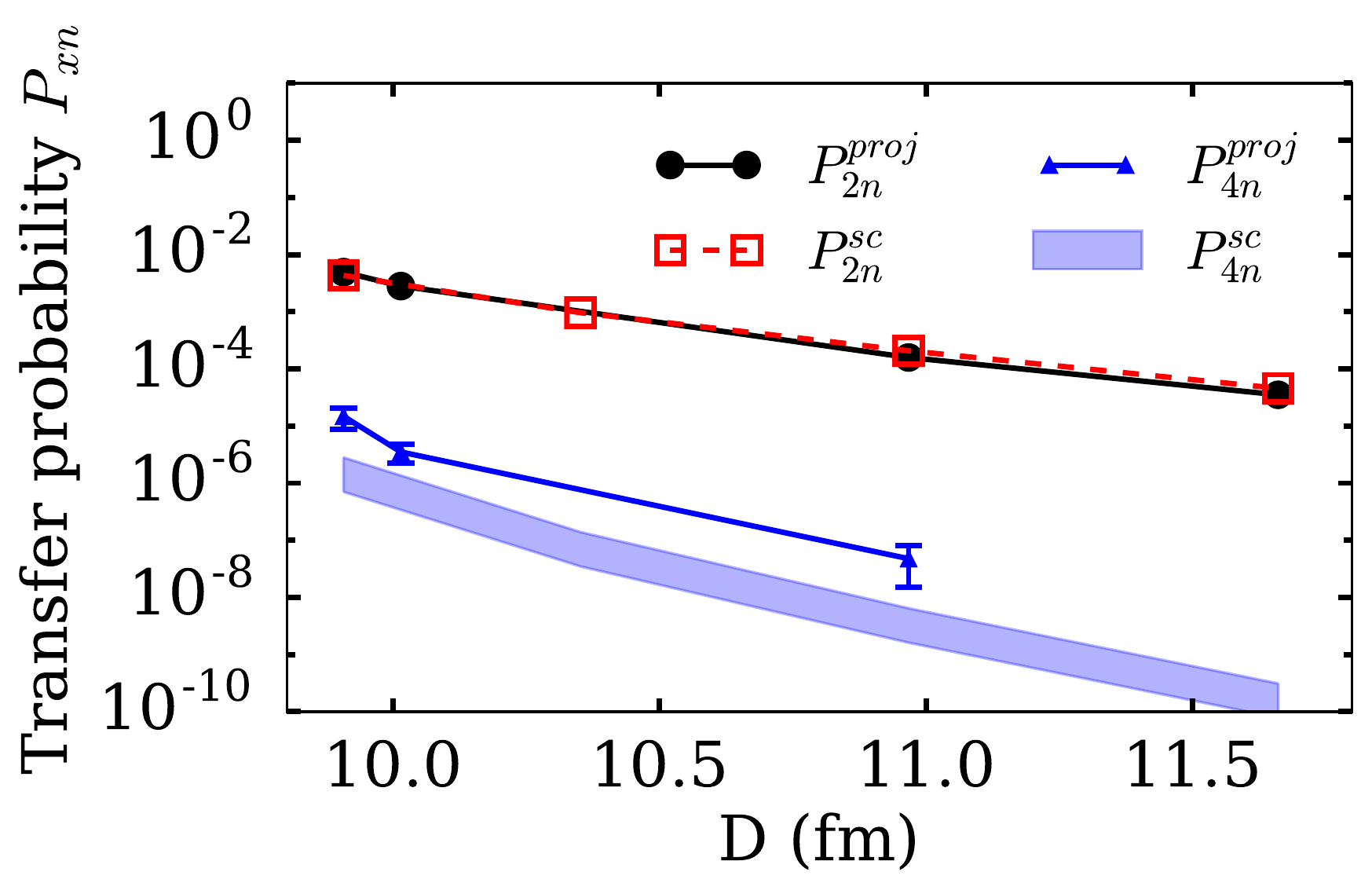}
 \caption{Probability to transfer one and two pairs in the symmetric central collision $^{20}$O + $^{20}$O as a function of the closest distance of
 approach $D$ during the reaction. The probabilities $P_{xn}^{proj}$ are computed with the approximate projection method of Ref.~\cite{Sca17b},whereas $P_{xn}^{sc}$ results from the PSC approach presented in this paper. The error bars associated to $P_{xn}^{proj}$ correspond to the fluctuations of the particle number after re-separation of the collision partners. They are too small to be visible for $P_{2n}^{proj}$. The colored band for $P^{sc}_{4n}$ spans the area obtained with the two prescriptions for the $W_k$ factors. These factors are computed using $n_A=n_B=2$ and 
 $\Omega_A=\Omega_B = 6$ (see text).}
 \label{fig:o20o20}
\end{figure}


\section{Applicability to reactions between non-identical superfluid systems}
\label{sec:gen_asym}
The major motivation for this work was the surprisingly large effects related to gauge angle in the collision between two identical superfluids observed in \cite{Mag17}. In this section, we give some hint on whether these effects are still present in the case of the contact between non-identical superfluids systems and if the PSC method is still applicable in this case.

\subsection{Schematic model for asymmetry reactions}

\subsubsection{Generalization of the PSC method for non-identical systems}
As a first step, we show how the PSC method modified to properly describe the pair transfer process in the schematic 
model considered previously when the two initial superfluids are different, i.e. when they have for instance different particle numbers, degeneracies, single-particle levels, \textit{etc}). 
Two major differences appear in this case: (i) the probability to transfer pairs comes not only from fluctuations but also from an average drift of the mean particle number transferred from one system to the other; (ii) the elementary probability to transfer a pair from $A$ to $B$ or from $B$ to $A$ are \textit{a priori} different, i.e. $p\neq q$ and in general  $P_{xn} \neq P_{-xn}$.

Starting from the same hypothesis (a-c), we generalize below the PSC technique to access the pair addition/removal probabilities.
In the perturbative regime, we  can still assume that only $P_{0n}$ , $P_{2n}$ and $P_{-2n}$ are dominating.
Denoting $\delta n_A = \bar{N}_A^\infty - N_A^0$ the average number of pairs transferred from $B$ to $A$, and using hypothesis (b), 
we can express the probabilities $P_{2n}$, and $P_{-2n}$. Indeed, we now have:
\begin{eqnarray}
\delta n_A & \simeq & 2 (P_{2n} - P_{-2n}),  \label{eq:dna}
\end{eqnarray}
while 
\begin{eqnarray}
\mu_2 & \simeq & 4 (P_{2n} + P_{-2n}). \label{eq:mu2}
\end{eqnarray}
Note that here, we used the fact that $P_{\pm 2n}^2 \ll P_{\pm 2n}$.  
Inverting these equations leads the simple expressions:
\begin{eqnarray}
\left\{
\begin{array}{c}
\displaystyle P_{2n} \simeq \frac{\mu_2 + 2 \delta n_A}{8} = p W_1 \\
\\
\displaystyle P_{-2n} \simeq \frac{\mu_2 - 2 \delta n_A}{8} = q W_{-1}
 \end{array}
\right. 
,
\label{eq:p2pm2}
\end{eqnarray}
that gives a straightforward method to extract the values of $p$ and $q$. 
The equation~(\ref{eq:recurence}) are now extended as  
\begin{eqnarray}
\left\{ 
\begin{array}{l}
P_{2 (k+1)n}  = p \frac{W_{k+1}}{W_{k}} P_{2kn},  \\
\\
P_{-2 (k+1)n}  = -q \frac{W_{-(k+1)}}{W_{k}} P_{-2kn} 
\end{array}
\right. .
\label{eq:recurence2}
\end{eqnarray}

It is worth mentioning that
the above expressions should be used with some care. Indeed, when the absolute value of the drift $|\delta n_A|$ 
increases and exceed $\mu_2/2$, one of the probabilities becomes negative which is unphysical. This directly stems 
from the breakdown of the hypothesis  (b). Indeed as $|\delta n_A|$ increases, on of the two particle transfer probabilities 
starts to decrease and becomes comparable to $P_{4n}$ and or $P_{-4n}$. This constraint on the applicability of equation (\ref{eq:p2pm2}) can be further quantified. Let us first assume that $\delta n_A>0$, which means that the transfer from $B$ to $A$ is dominant and 
$P_{2n} > P_{-2n}$. Then, the condition $P_{-2n} \gg P_{4n}$ gives the condition:
\begin{eqnarray}
\left( \frac{8 W^2_1}{W_2} \right) \frac{\left[ \mu_2 - 2 \delta n_A\right] }{\left[\mu_2 + 2 \delta n_A\right]^2} \gg 1 .
\end{eqnarray}
Equivalently for the case $\delta n_A < 0$, which corresponds to the case where the transfer from $A$ to $B$ dominates, the condition 
becomes:
\begin{eqnarray}
\left( \frac{8 W^2_{-1}}{W_{-2}} \right) \frac{\left[ \mu_2 + 2 \delta n_A\right] }{\left[\mu_2 - 2 \delta n_A\right]^2} \gg 1 .
\end{eqnarray}

If these conditions are not meet, i.e. if the drift becomes too high compared to the second moment, one of the 
two-particle transfer probability dominates the other. Then, Eqs. (\ref{eq:dna}) and (\ref{eq:mu2}) simplify and we have: 
\begin{eqnarray}
\mu_2 & \simeq & 2 |\delta n_A| \simeq 4 P_{\eta 2  n}, \label{eq:mu2pasym}
\end{eqnarray} 
where $\eta $ is the sign of the $\delta n_A$. 
The moments of order one and two contain a redundant information which characterizes only the dominant branch (addition or removal). In practice, the probability $p$ (or $q$) associated to this branch may still be obtained by matching the first moment with $2P_{\eta2  n}$.
It is finally interesting to mention, that for large drift we obtain $\mu_2  \propto   |\delta n_A| $ that is similar to the results obtained in the nucleon exchange model where the transfer is driven by randomness \cite{Ran80,Fel87,Was09}.

\subsubsection{Benchmark with the exact second moment}

To benchmark this approach, we follow the same methodology as for the symmetric case and first compute the probability distribution $P_{xn}$ from the exact average drift and second moment.
In the present study, we consider three different types of asymmetric reactions: (I) the case of degenerate system with a different initial number of particles in each system $N_A^0=4$, $N_B^0=8$; (II) the same situation with an additional asymmetry coming from a shift in the single-particle energy between the two systems, namely $\Delta e = \epsilon_B -\epsilon_A = g$; (III) the case of two non-degenerated systems with equidistant single-particle level spacing $\Delta \varepsilon / g = 1$ in each system and different initial number of particles in $A$ and $B$. In all cases, we assume $\Omega_A=\Omega_B= 6$. Note that we also tested situations where $\Omega_A \neq \Omega_B$ leading essentially to the same conclusions.   
 

\begin{figure}[h!]
 \includegraphics[width=0.9\linewidth]{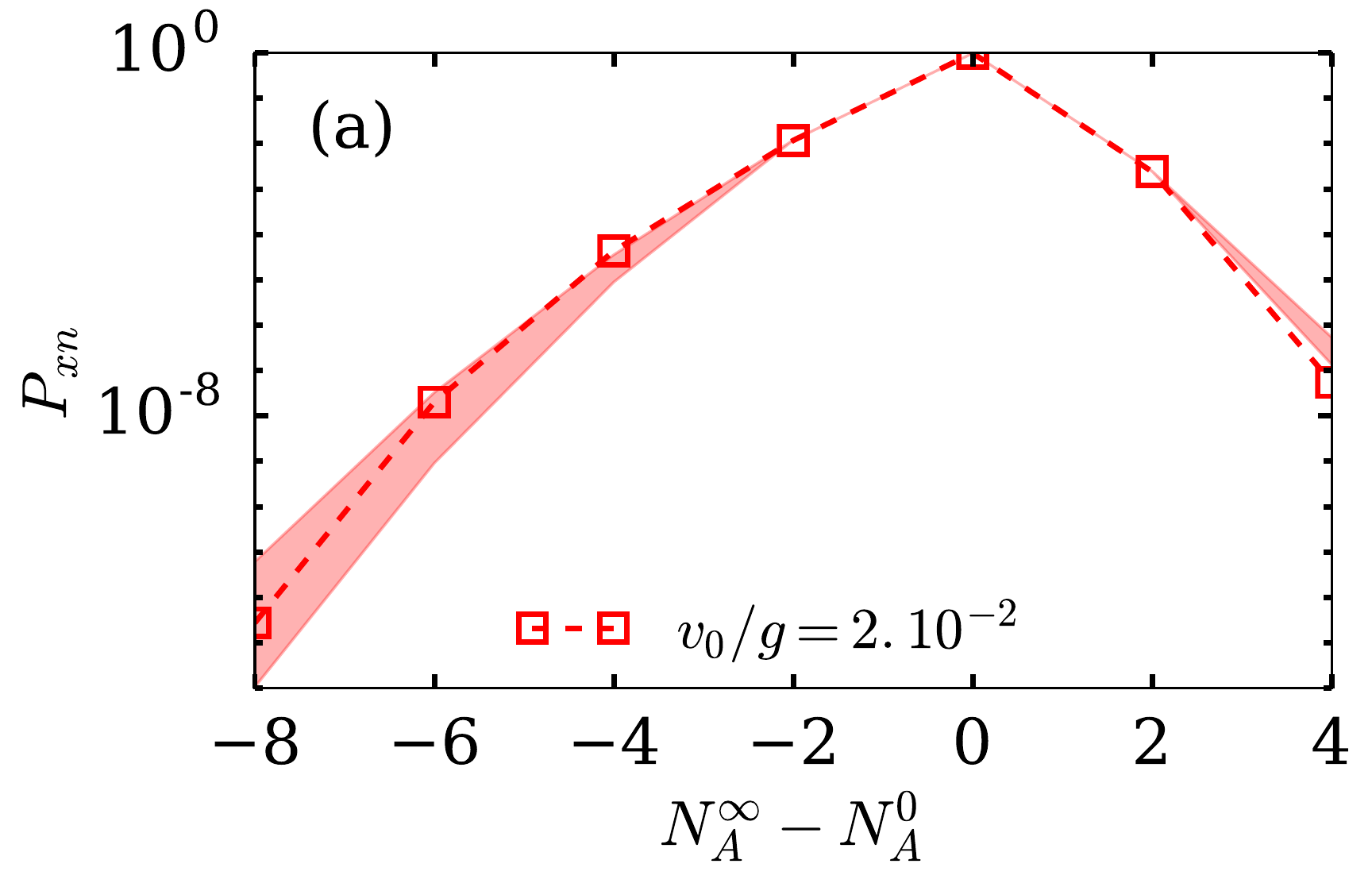}
 \includegraphics[width=0.9\linewidth]{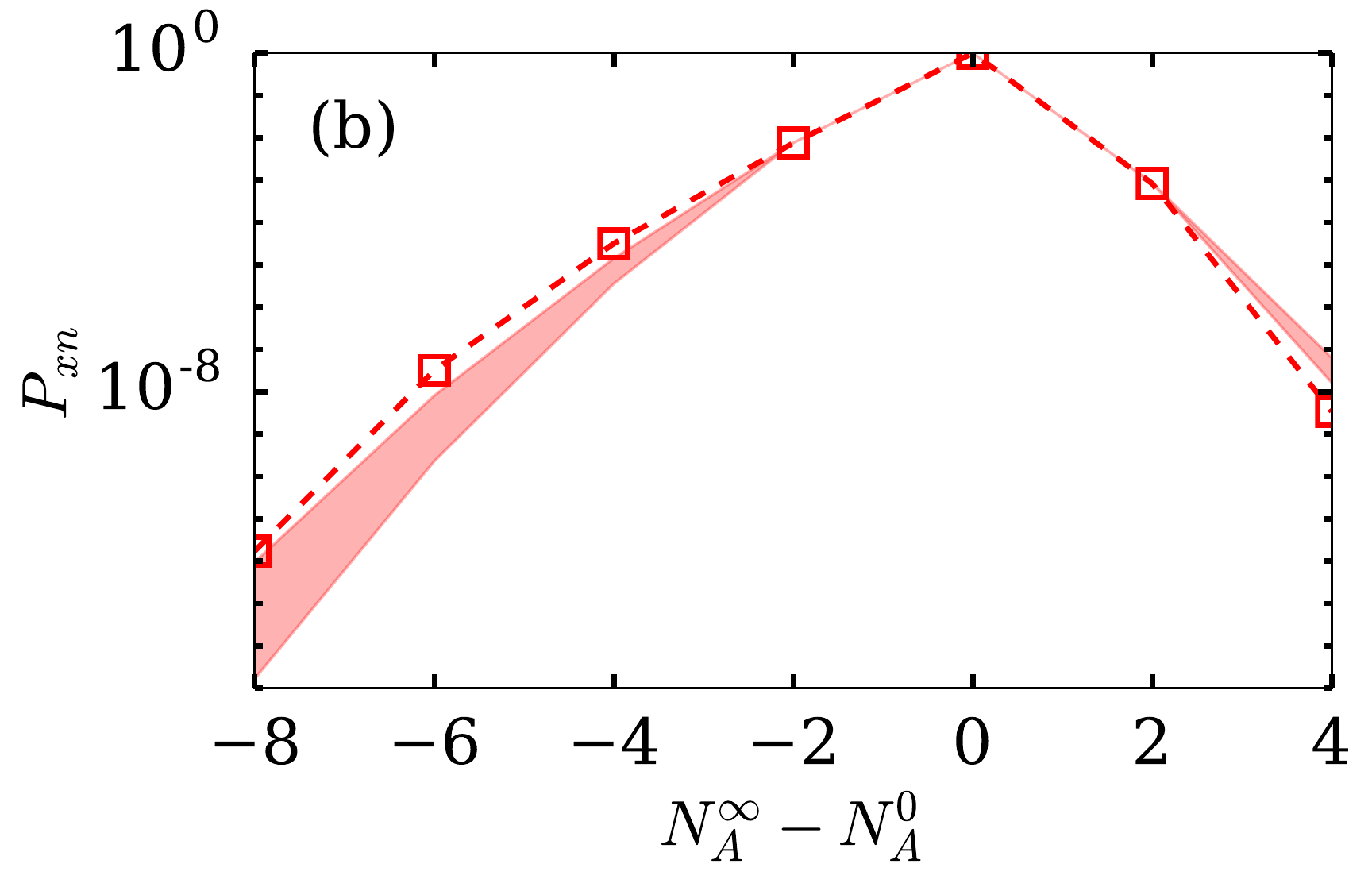}
 \includegraphics[width=0.9\linewidth]{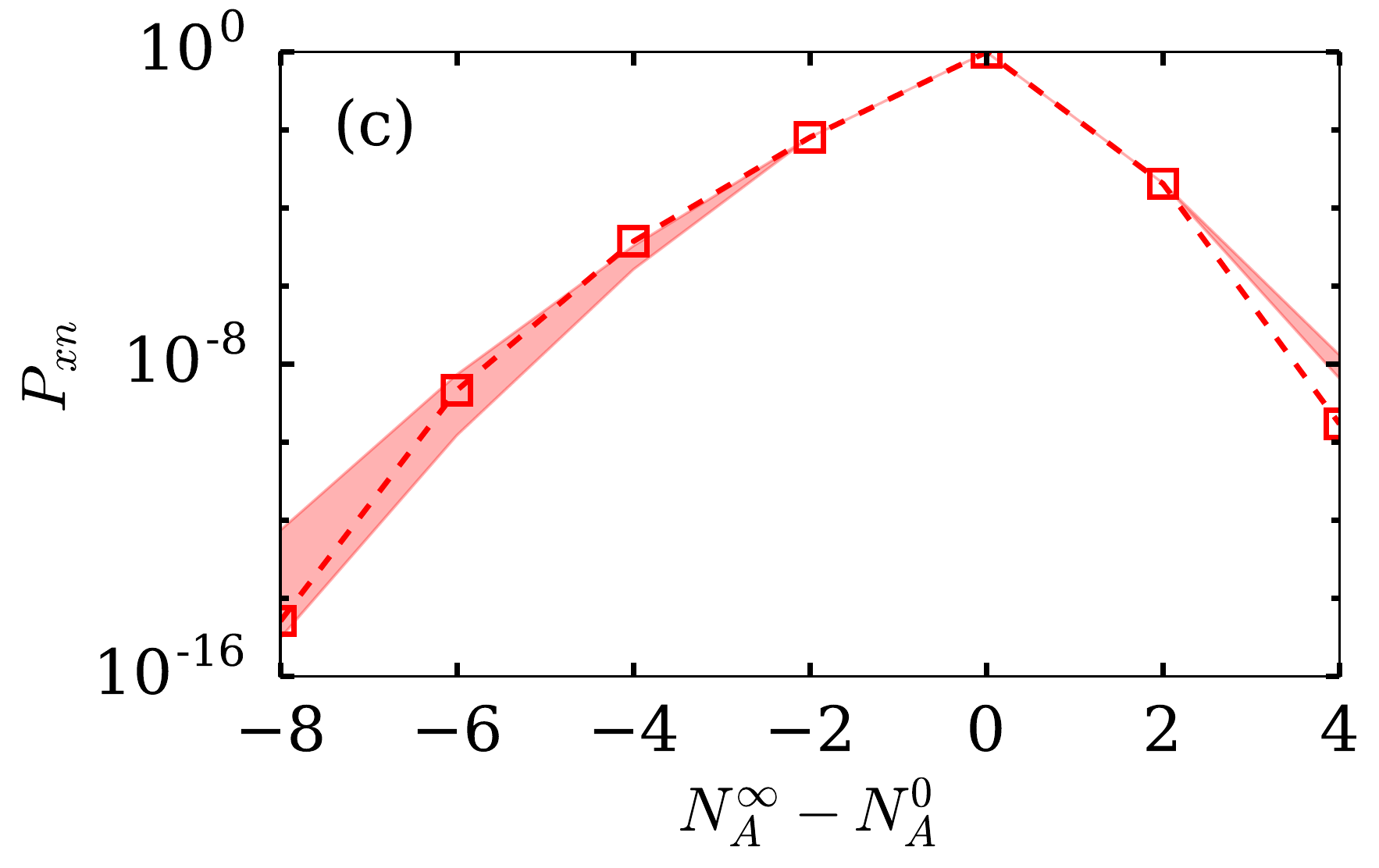}
 \caption{Same as Fig. \ref{fig:distrib_exact} for three cases of asymmetric reactions. The exact probabilities are shown with dashed lines and systematically compared to the combinatorial results using the exact values of $\bar{n}$ and $\mu_2$  (colored bands). The different panels (a), (b) and (c) correspond respectively to the asymmetric case (I), (II) and (III).}
 \label{fig:distribasym}
\end{figure}
In Fig.~\ref{fig:distribasym} the pair transfer probabilities obtained for the three asymmetric reactions using the PSC approach  with the exact $\delta n_A$ and second moment. These probabilities are compared to the exact ones. We end up essentially to the same conclusions as in the symmetric case. We systematically see that the probability to transfer one or two pairs are reasonably well reproduced, while as the number of transferred pair increases, more and more deviation appears with respect to the exact distribution. Overall, the shape of the distribution still matches correctly the exact one.

\subsubsection{Critical discussion on the semi-classical moments in collisions between non-identical systems}

Note also that for the case of asymmetric collisions that is considered here, there is no more strong argument to apply the same scaling for $g_A,g_B$ and $v(t)$ to compare with the exact solution. For the sake of simplicity, we kept the same procedure as for the symmetric case. 

For the three asymmetric cases considered, we compute the semi-classical moments $\delta n_A^{\rm sc}$ and $\mu^{\rm sc}_2$ from the average over different TDHFB trajectories. When going from symmetric case to asymmetric case, we did not found any systematic arguments to obtain the scaling on the coupling constant discussed in section~\ref{sec:crit}. 
for the sake of simplicity, we therefore adopted the same scaling procedure as in the symmetric case and found $\alpha=1.2$ for cases (I) and (II) and $\alpha=1.289$ for case (III). 

The figure~\ref{fig:cmuv0ratioasym} compares the semi-classical estimation of the two first moments with the exact one. In the three cases, this procedure yields moments that are proportional to their exact counterparts in the perturbative region. The semi-classical drift lies within 20\% of the exact drift in this regime, whereas the second moment is systematically underestimated, $\mu^{\rm sc}_2\sim 0.5-0.7\mu_2^{ex}$.
The absence of clear procedure to generalize the scaling technique strongly bias the comparison between the exact and the semi-classical approach combined with the combinatorial technique. From these results, it is not clear whether the underestimation of the second order moment comes essentially from the arbitrary rescaling procedure or an intrinsic feature of the semi-classical phase-space method. 
This question is crucial as such an underestimation in a realistic case would severely jeopardize the method proposed for asymmetric reactions. For example, in the model case (II) and (III) we find that whatever the values of $p$ and $q$, no distribution represented by Eq.~(\ref{eq:pksimp}) could reproduce the moments estimated from the semi-classical average.
\begin{figure}[h!]
 \includegraphics[width=0.9\linewidth]{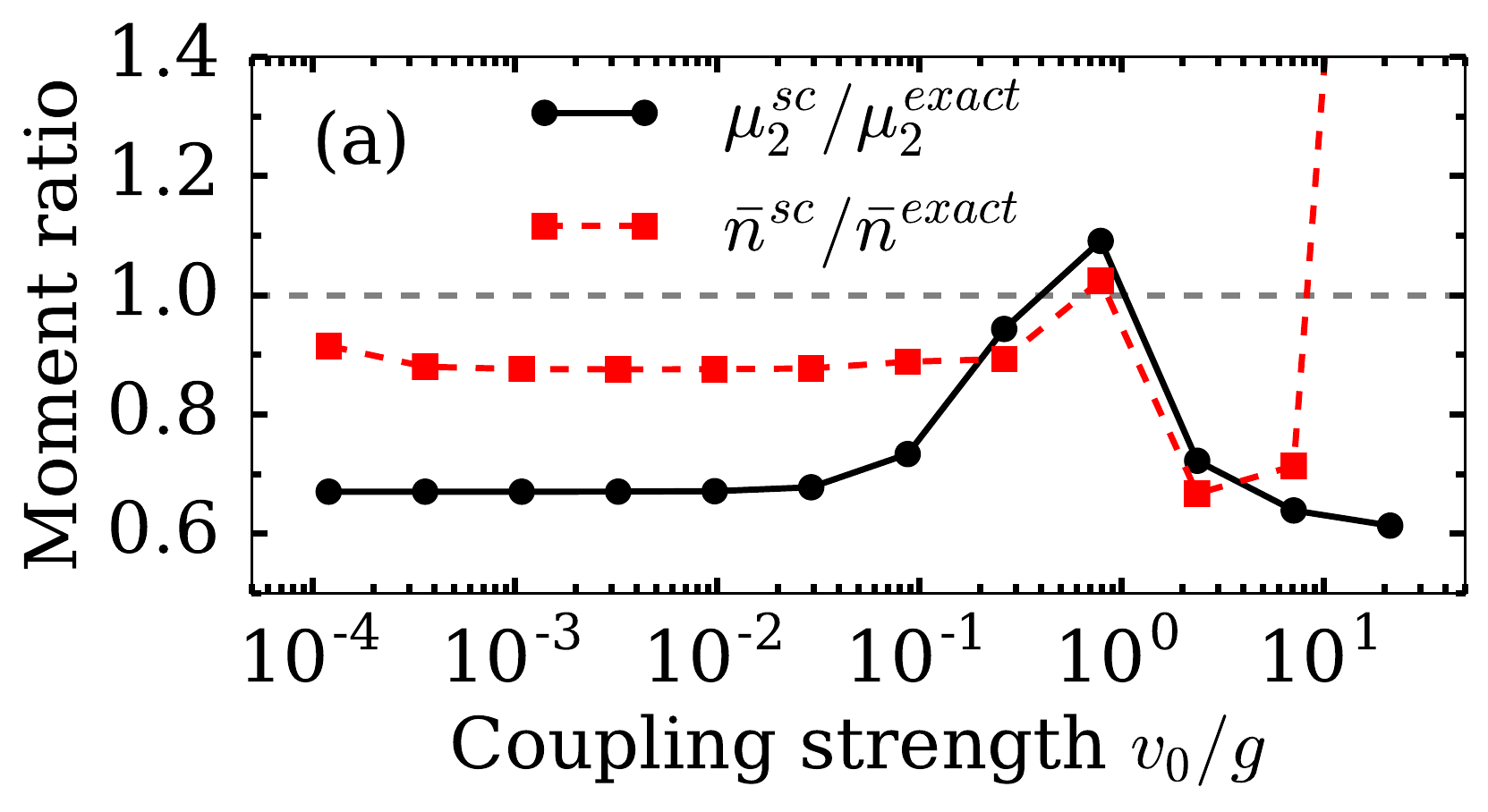}
 \includegraphics[width=0.9\linewidth]{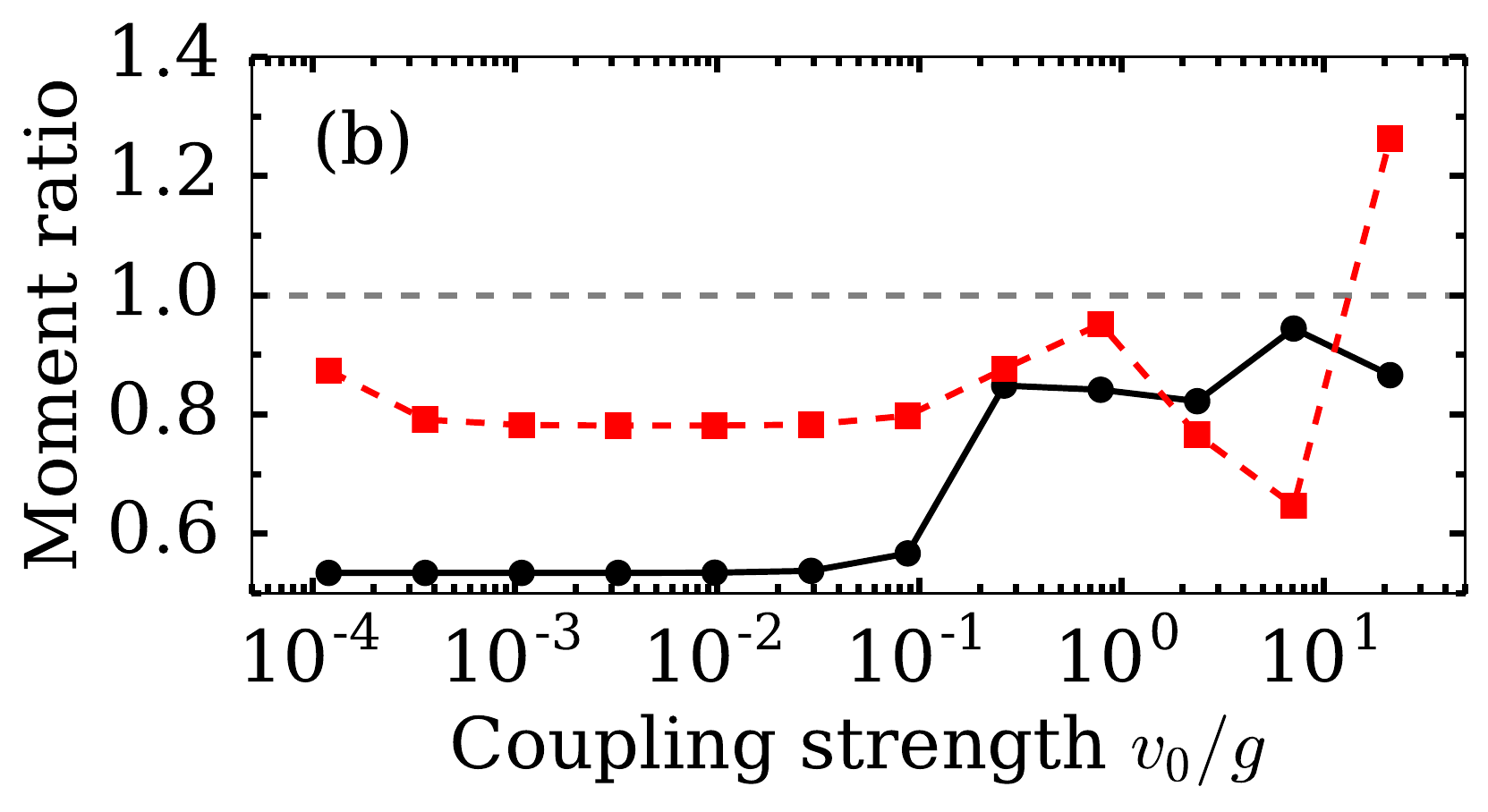}
 \includegraphics[width=0.9\linewidth]{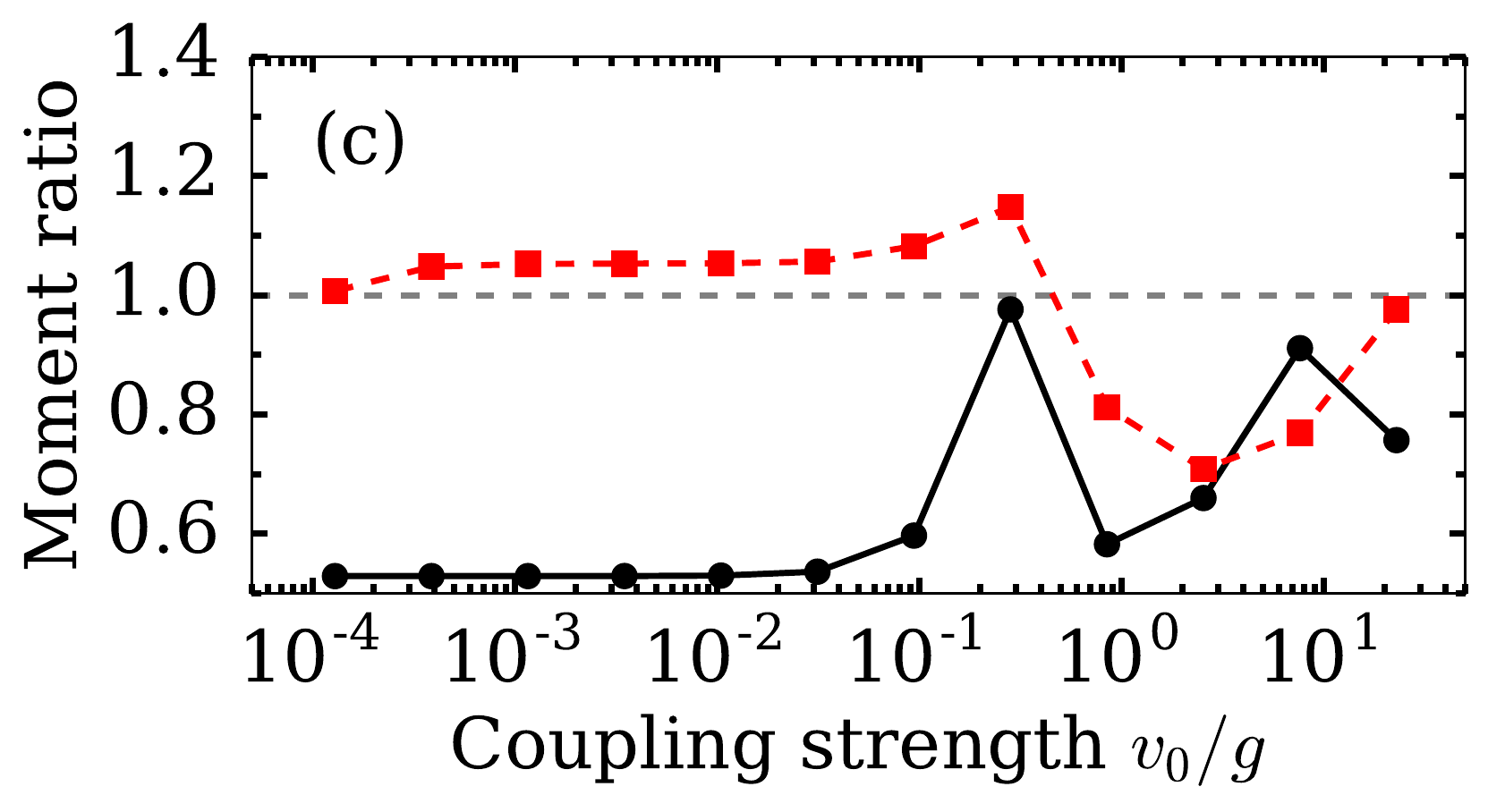}
 \caption{First and second order moments  obtained with the semi-classical average normalized to their exact counterparts as a function of the coupling strength. The different panels (a), (b) and (c) correspond respectively to the asymmetric case (I), (II) and (III).}
 \label{fig:cmuv0ratioasym}
\end{figure}

To conclude, a generalization of the PSC method to the case of non-identical superfluid systems is technically possible. Its application to our toy model shows that in the case where the transfer is dominated by an average drift of pairs only a part of the distribution could be recovered. In addition, it is not clear from this study whether the low order moments of the distribution can be correctly estimated within the phase-space averaging procedure. To answer this remaining question, an application of the PSC method to a realistic collision between non-identical systems is required.
%
%

\subsection{Application to $^{14}$O + $^{20}$O reaction below the Coulomb barrier}

Using the semi-classical average over the initial relative gauge angle, we computed the semi-classical drift and second order moment of the pair transfer distribution for the asymmetric reaction $^{14}$O + $^{20}$O. 
The calculations were repeated at three different energies below the Coulomb barrier and the table~\ref{tab:o14o20} summarizes the results. 
For the sake of completeness, we also give the results obtained by the projection method proposed in Ref. \cite{Sca17}.
In the case of non-identical nuclei, 
almost no influence of the initial gauge-angle on the particle transfer process is found leading to a small, almost zero, value of 
$\mu_2^{sc}$ from which we deduce very small values for the pair transfer using Eq. (\ref{eq:mu2pasym}). This is at variance with both the toy model for a symmetric reaction and the $^{20}$O + $^{20}$O application. This is also different from the toy model used previously for interacting non-identical superfluids. 
The difference with the toy model can be directly traced back to the fact that one-body components treating single-particle tunneling 
has not been considered in the toy model. This process seems to dominate for the asymmetric collisions considered here washing out any 
significant effects of gauge-angle. 

As a matter of fact, this is not surprising because the transfer process in asymmetric systems is known to be mainly dominated by the fast
$N/Z$ equilibration, due to a fast equilibration of the chemical potentials. This process is already accounted for by the mean-field hamiltonian and is not connected to superfluidity. It is worth mentioning that mean-field alone (with or without pairing) can describe the mean drift but not the fluctuations around the mean drift.  
 
The absence of gauge-angle influence obviously leads to a failure of the PSC approach. Indeed, a prerequisite of the success of the approach 
presented here is that the U(1) symmetry breaking dominates the physical process under interest.  

Our conclusions are twofold. First, the PSC method applied to nuclear reactions can only be successful for symmetric collisions between superfluid nuclei to avoid at maximum the pollution from pure one-body effects. Second, for the same reason, experiments involving symmetric 
reactions between mid-shell spherical nuclei is the best test-bench to probe any significant gauge-angle effects, if any.  
\begin{table}[ht!]
\begin{tabular}{ccccc}
\hline
$E_{cm}$ & $\delta n_{A}$ & $P_{2n}^{sc}$  & $P_{1n}^{proj}$ & $P_{2n}^{proj}$ \\
(MeV)& & =$\mu^{sc}_2/4$  & & \\
\hline
7.903 & 0.28$\times10^{-3}$   & 0.04$\times 10^{-9}$ & 0.28$\times 10^{-3}$   & 0.05$\times 10^{-3}$ \\
8.903 & 7.93$\times10^{-3}$   & 0.14$\times 10^{-9}$ & 3.31$\times 10^{-3}$   & 1.23$\times 10^{-3}$ \\
9.403 & 40.11$\times10^{-3}$  & 0.48$\times 10^{-9}$ & 13.04$\times 10^{-3}$ & 7.11$\times 10^{-3}$  \\
\hline
\end{tabular}
\caption{Average drift along with fluctuation of the number of transfered particles (and pair transfer probabilities $P_{2n}^{sc}$) obtained within a phase-space averaging for the $^{14}O + ^{20}O$ reaction at different energies in the center of mass ($E_{cm}$). The probabilities to transfer one $P_{1n}^{proj}$ and two $P_{2n}^{proj}$ particles obtained from an approximated projection technique are also provided for comparison.}
\label{tab:o14o20}
\end{table}

\section{Conclusion}

In this work, we analyze in detail the pair transfer between two identical superfluid systems. We use a simple model where the two systems governed by a pairing Hamiltonian only interact with each other for a short finite time. This model mimics the transfer of pairs during a heavy-ion reaction in the presence of pairing and can be solved exactly for small systems. The possibility to describe accurately the pair transfer by combining the TDHFB framework with some average over the relative gauge angle is discussed. We show in particular that a brute force semi-classical average over a distribution of gauge angles can only partially describe the final distribution of particles in each subsystem. While the second moment of this distribution is reasonably described, higher order moments significantly differ from the exact results. 

Based on this observation, a method is proposed to obtain the probability to transfer multiple pairs in the perturbative regime. This approach supplements the phase-space average with a combinatorial technique to infer the probabilities to transfer more than one pair. 
The PSC method is benchmarked with respect to the schematic model and then successfully applied to the head-on collision of two $^{20}$O. For this realistic case, the results are systematically compared with the projection method proposed in Ref. \cite{Sca17b}. Despite the fact that the PSC method is technically much simpler than the projection method, both techniques lead to similar two particle transfer probabilities. The PSC that was shown to be effective in the schematic model to reproduce multi-particle transfer leads in general to multiple pairs transfer probabilities that are much smaller than the projection approach. 

Finally, the applicability of the PSC method to the case of asymmetric collision is discussed. We show in particular that in its current state, the method fails to describe the transfer in the asymmetric reaction $^{14}$O+$^{20}$O. The main reason is that the fluctuations associated with the relative gauge angle in this reactions are not the main driver of the transfer. 

In the future, it would be interesting to compare the different approaches to describe multiple pair transfer 
to high-precision experiments. In recent years, several experiments have been performed for energies well below the Coulomb barrier~\cite{Cor11,Mon14,Mon16,Raf16} where the perturbative regime is relevant. However, these experiments  involve targets different from the projectile where many effects other than pairing play a role in the transfer. We believe that the best situation to compare theory and experience and to unambiguously probe relative gauge-angle effects in multiple particles transfer, would be to consider symmetric collisions between medium mass spherical nuclei, like $^{108-120}$Sn, where the last occupied level is close to half-filling. 

\appendix

\section{Exact evolution for the symmetric degenerate case}
\label{ap:exa}

In this section, we show that the combinatorial approach used in the present work can be motivated in some way by the exact case in the perturbative regime. We consider here that the two systems $A$ and $B$ are fully degenerated and governed by a pure pairing Hamiltonian. We follow Ref. \cite{Die71} and use the compact notation $H_0 = H_A+H_B$. We assume that the system is described initially by the state $| n_A, n_B \rangle= | n_A \rangle \otimes | n_B \rangle$ where $| n_A \rangle$ and  $| n_B \rangle$ are respectively the seniority zero ground states of $A$ and $B$. The symbol $n_{A/B} = N_{A/B}/2$ denotes here the number of pairs. Since, the Hamiltonian conserves the seniority and the total number of particles, in the simple symmetric degenerate case (with initial half filling), one can introduce the set of states 
\begin{eqnarray}
| k \rangle \equiv | n_A + k, n_B - k \rangle, ~~-n_A \le k \le +n_B \nonumber,
\end{eqnarray}
associated to the unperturbed energy $E_n$ and decompose the time-dependent state $|\Psi(t)\rangle$ as:
\begin{eqnarray}
|\Psi(t)\rangle &=& \sum_k c_k(t) | k \rangle,
\end{eqnarray}
with $|\Psi(t_0)\rangle = | 0 \rangle$. Introducing the set of parameters $b_k(t) = c_k(t) e^{+iE_k t/\hbar} $, the coupled equations 
on the $b_k$ components writes:
\begin{eqnarray}
i\hbar \frac{d}{dt} b_l(t) & = 
& v(t) \sum_k b_k(t)  e^{i \omega_{kl}t}\langle l |(C^\dagger+C)| k \rangle \nonumber,
\end{eqnarray}
with $\omega_{kl}  =   (E_l -E_k )/\hbar$ and where we have introduced the notation $C^\dagger = \sum_{\alpha \beta} a^\dagger_{ \alpha} a^\dagger_{\bar \alpha} b_{\bar \beta} b_\beta$. 
One can then introduce two matrices $D^{+}$ and $D^{-}$ with components
\begin{eqnarray}
\left\{
\begin{array} {c}
\displaystyle
D^{+}_{lk} = \langle l |C^\dagger| k \rangle = \delta_{l,k+1} D^{+}_{k},  \\
\\
\displaystyle D^{-}_{lk} = \langle l |C| k \rangle =  \delta_{l,k-1} D^{-}_{k}. 
\end{array}
\right.
\label{eq:dpm}
\end{eqnarray} 
Note that $D^+$ (respectively $D^-$) only connects the state $|k\rangle$ with the state $| k+1 \rangle$ (resp. $|k-1 \rangle$).
We first assume that the transition frequencies are constant with:
\begin{eqnarray}
\omega_{kl} = \pm \omega. \label{eq:omegacte}
\end{eqnarray}
On top of that, we assume that the perturbation is time symmetric and take for convenience $t_\infty = - t_0$.
From this, we obtain a compact 
expression for the $b$ vector:
\begin{eqnarray}
b(t_\infty) &=& \exp(2 i z  [D^+ +D^-] ) \, b(t_0),
\end{eqnarray} 
with 
\begin{eqnarray}
z = - \frac{1}{\hbar}\int_{t_0}^{t_\infty} ds\,  v(s) \cos(\omega s). 
\end{eqnarray}
The final probability to have $k$ pairs transferred, denoted again by $P_{2kn}$ from the initial state $| 0\rangle$ is then given by:
\begin{align}
\label{eq:probexact}
P_{2kn} =&   |b(t_\infty) |^2 \nonumber \\
=& \left| \sum_{J=0}^{+\infty} \frac{(2i z)^J}{J!} \left[ \left(D^+ +D^- \right)^J \right]b(t_0) \right|^2 \\
=& \left| \sum_{J=0}^{+\infty} \frac{(2i z)^J}{J!} \left[  \sum_{I=0}^J C^I_{J} (D^+)^I (D^-)^{(J-I)}   \right] b(t_0) \right|^2  \nonumber .
\end{align}  
In general, the transfer probabilities result from a rather complicated interference effect between different 
paths depicted in Fig. \ref{fig:interference}. In this double sum, all terms with $I-(J-I)= 2I -J =k$ will contribute 
to the probability $P_{2kn}$.
However, noting that $z \propto v_0$, the first term feeding the state $| k\rangle$ in Fig. \ref{fig:interference} will dominates the probability in the perturbative regime. This term corresponds always to the lowest or highest branch in this figure. 
\begin{figure}[h!]
 \includegraphics[width=0.9\linewidth]{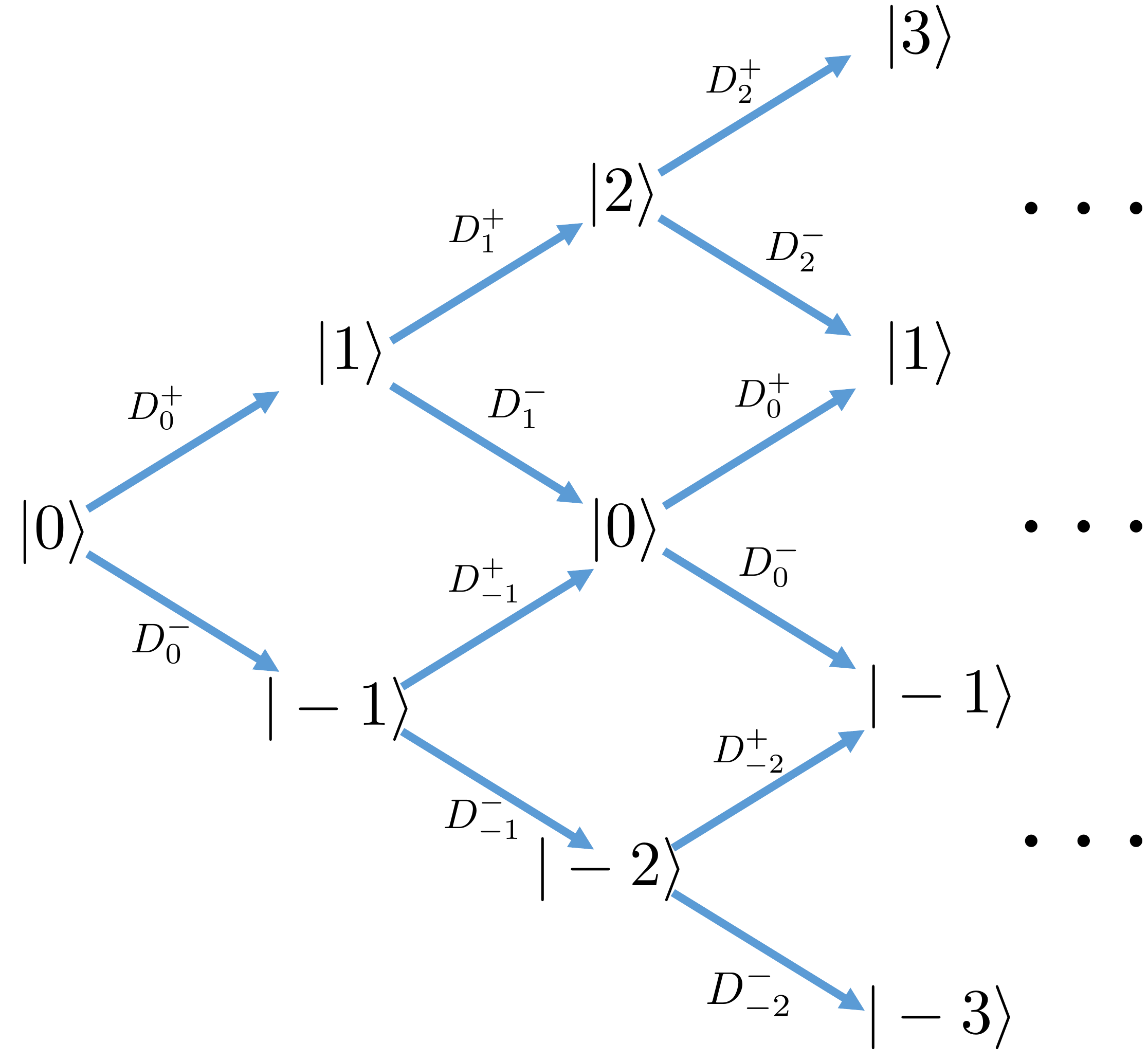}
 \caption{Schematic view of the different contributions to the population of a state $|k\rangle$ in equation  (\ref{eq:probexact}).}
 \label{fig:interference}
\end{figure}
Neglecting all other terms in the expansion yields the simplified expressions:
\begin{eqnarray}
P_{2kn}& \simeq&  \frac{|2 z|^{2k}}{(k!)^2}  \left| D^+_0 \right|^2 \cdots  \left| D^+_{k-1} \right|^2 \nonumber \\
P_{-2kn}& \simeq&  \frac{|2 z|^{2k}}{(k!)^2}  \left| D^-_0 \right|^2 \cdots  \left| D^-_{-(k-1)} \right|^2 \nonumber .
\end{eqnarray}
These formula induce the recurrence relations:
\begin{eqnarray}
P_{2(k+1)n} & = & |2 z|^2 \frac{ \left| D^+_{k} \right|^2}{(k+1)^2}   P_{2kn}, \nonumber \\
P_{-2(k+1)n} & = &   |2 z|^2 \frac{ \left| D^-_{k} \right|^2}{(k+1)^2} P_{-2kn}.  \nonumber 
\end{eqnarray}
The matrix elements of $D^+, D^-$ read~\cite{Die70}
\footnote{As a side remark, we note that if the exact state with particles $N_A,N_B$ is systematically replaced by a BCS/HFB quasi-particle state with average particle number $N_A$, $N_B$, we obtain 
\begin{eqnarray}
|D^+_0|^2 & = & \left( \frac{\Delta_A(N_A)\Delta_B (N_B)}{g_Ag_B}  \right)^2 \nonumber \\
&=& n_An_B (\Omega_A - n_A) (\Omega_B - n_B)  , \nonumber
\end{eqnarray} 
where $\Delta^{N_A}_A$ and $\Delta^{N_B}_B$. Similarly, we have:
\begin{eqnarray}
|D^+_k|^2 & = & (n_A +k) (\Omega_A - n_A-k)  (n_B - k ) (\Omega_B - n_B + k) , \nonumber \\
|D^+_{-k}|^2 & = & (n_A -k) (\Omega_A - n_A+k)  (n_B + k ) (\Omega_B - n_B - k) , \nonumber
\end{eqnarray} 
that is rather close to the exact formula.}
:
\begin{eqnarray}
|D^+_k|^2 & = & (\Omega_A  - n_A-k)(n_A+k+1) \nonumber \\
&&\times (n_B-k) (\Omega_B - n_B +k+1), \label{eq:dpk} \\
|D^-_{-k}|^2 & = & (\Omega_B  - n_B-k)(n_B+k+1) \nonumber \\
&&\times  (n_A-k)(\Omega_A - n_A + k +1) , \label{eq:dmk}
\end{eqnarray}
$n_A$ and $n_B$ being the initial number of pairs 
in $A$ and $B$ respectively. We finally deduce the compact expression:
 \begin{eqnarray}
P_{2kn}&=&  |2 z|^{2k} C^k_{\Omega_A - n_A} C^k_{n_B}   \frac{(n_A + k)! }{n_A!}  \frac{ (\Omega_B - n_B + k )! }{(\Omega_B - n_B)!}  ,
\nonumber \\
P_{-2kn}&=& |2 z|^{2k} C^k_{\Omega_B - n_B} C^k_{n_A}   \frac{(n_B + k)! }{n_B!}  \frac{ (\Omega_A - n_A + k )! }{(\Omega_A - n_A)!} . \nonumber
\end{eqnarray}
In the above expressions, we recognize the combinatorial factor that has been introduced  in the main text. 
Choosing $p \simeq |2 z|^{2}$, we obtain indeed
\begin{eqnarray}
W_k   =  
C^k_{\Omega_A - n_A} C^k_{n_B}   \frac{(n_A + k)! }{n_A!}  \frac{ (\Omega_B - n_B + k )! }{(\Omega_B - n_B)!} , \label{eq:appwk} \\
W_{-k}  = 
C^k_{\Omega_B - n_B} C^k_{n_A}   \frac{(n_B + k)! }{n_B!}  \frac{ (\Omega_A - n_A + k )! }{(\Omega_A - n_A)!} , 
\label{eq:approx1}
\end{eqnarray} 
for the factors $W_{k/-k}$ involved in equation (\ref{eq:simplepk}).

In Fig.~\ref{fig:distrib_n12_sym_gkk0_app}, the method proposed in section \ref{sec:transferproba} to obtain the transferred probabilities with two prescriptions for the $W_k$ is tested against the exact results. In this figure, we also display the results obtained from the coupled equation enforcing that all transition frequencies are constant, as it was supposed in the derivation here.
By construction, the results from the combinatorial method using equation (\ref{eq:wkcst}) for the factors $W_k$ matches the results from the coupled equation resolution with constant transition frequencies. On the other hand, this method  tends to overestimate the transfer probabilities compared to the exact case especially when the number of pairs transferred increases. We conclude from this that (i) the approximation (\ref{eq:omegacte}) is too crude 
for the present situation; (ii) there is a rather close connection between the combinatorial factor and 
the set of values of transition frequencies at play during multiple pair transfer. 
\begin{figure}[ht!]
 \includegraphics[width=0.9\linewidth]{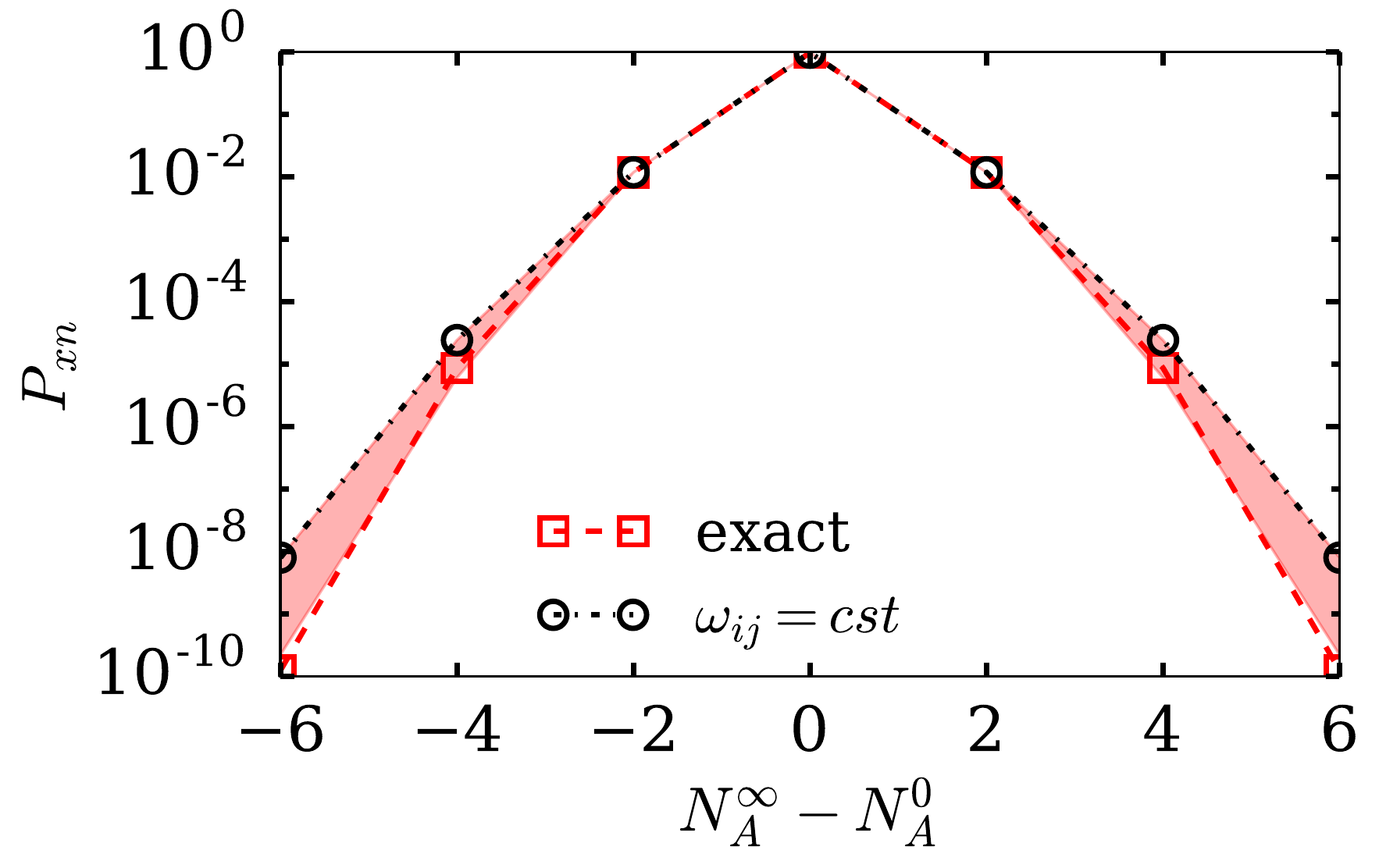}
 \caption{Asymptotic probabilities to transfer one or several pairs as a function of  the quantity
 $N_A^{\infty}-N_A^{0}$ obtained from a symmetric degenerate situation with $\Omega_A=\Omega_B=N_A^0=N_B^0=$ 6. The colored bands span the area obtained with Eq.~(\ref{eq:simplepk}) assuming Eq. (\ref{eq:appwk}) (upper limit) or the same expression divided by $(k!)^2$ 
 (lower limit). The distributions are obtained by using the exact second moment $\mu^{\rm sc}_2$ to determine $p=q$.  Approximate probabilities are compared with the exact values (red dashed line) and the leading order of the perturbative approach computed with the assumption $\omega_{ij}=cst$ (black dotted line). All calculations are performed for the coupling strength $v_0/g=2.10^{-2}$.}
 \label{fig:distrib_n12_sym_gkk0_app}
\end{figure}
Starting from a situation of a single degenerate shell with initial half-filling and neglecting pairing correlations, the addition or removal of 2 particles to the shell will indeed lead to $\omega_{kl} = \pm \omega = \pm 2 \varepsilon$ where $ \varepsilon$ is the single-particle energy in the shell. 
However, when pairing plays a role, the energies are distorted by correlations. In particular, starting from the exact energies of the degenerate system with initial half-filling, it can be shown that the energies verifies $\omega_{k, k+1} \simeq \pm  k \omega$ and therefore increases when the number of pairs increases. The combinatorial factor should take into account this aspect in some way. 

To obtain a more realistic expression of $W_{k,-k}$ for superfluid system, we start back from the time dependent perturbation theory. Without any assumption on the transition frequencies, we can show that the leading order contribution to $b_k$ with $k>0$ writes:
\begin{widetext}
\begin{eqnarray}
b_k(t_\infty) = \hbar^{-k} e^{-i(\omega_k t_\infty - \omega_0 t_0 )}
\int^{t_\infty}_{\tau_{k-1}} d\tau_k \int_{\tau_{k-2}}^{\tau_k} d\tau_{k-1} \cdots \int^{\tau_2}_{t_0} d\tau_1 \,
v(\tau_k) \cdots v(\tau_1) e^{-i\omega_{k,k-1}\tau_k}\cdots e^{-i\omega_{1,0}\tau_1} D^+_{k-1} \cdots D^+_0 . \nonumber
\end{eqnarray}
\end{widetext}
To further progress, we make the simplifying assumption that:
 \begin{eqnarray}
 \int^{t_\infty}_{\tau_{k-1}} d\tau_k \int_{\tau_{k-2}}^{\tau_k} d\tau_{k-1} \cdots 
\rightarrow \frac{1}{k!}
\int^{t_\infty}_{t_0} d\tau_k\int_{t_0}^{t_\infty} d\tau_{k-1} \cdots , \nonumber
\end{eqnarray}
where the factor $1/k!$ accounts properly for the change of volume of integration in the time hyper-space $(\tau_1, \cdots, \tau_k)$. We further simplify the interaction and assume $v(t) = v_0 \Theta (\tau_c/2 - |t|)$ where $\tau_c$ is the interaction time and $\Theta$ is the Heavyside function. With this, we can explicitly perform the time integration and obtain:
\begin{eqnarray}
|b_k(t_\infty)|  =  \frac{1}{k!} \left( \frac{2 v_0}{\hbar}\right)^k \prod_{l=1}^{k} \left[ \frac{ \sin\left( \omega_{l,l-1}  \tau_c/2 \right)}{\omega_{l,l-1} } \right]
  D^+_{k-1} \cdots D^+_0 , \nonumber
\end{eqnarray}
We then see that the probability to transfer $k$ pairs can be rewritten as a product:
\begin{eqnarray}
P_{2kn} =  {\cal P}_{+1} \cdots {\cal P}_{+k},
\end{eqnarray}
with:
\begin{eqnarray}
{\cal P}_{+k} &=& \left( \frac{2v_0}{ k \hbar} \frac{  \sin\left( \omega_{k,k-1}  \tau_c/2 \right)}{\omega_{k,k-1} }  \right)^2 |D^+_{k-1}|^2 
\label{eq:pksimp}
\end{eqnarray}
This probability is in general rather complicated since it contains the information on the initial and final system after transfer as well as the information on the time-dependent interaction. 

In the case where all the transition frequencies are equal to $\omega$, we recover the previous expression provided that 
\begin{eqnarray}
|2z| & = & \frac{|2 v_0|}{\hbar}  \left| \frac{ \sin\left( \omega  \tau_c /2 \right)}{\omega } \right|. \nonumber 
\end{eqnarray}
In a different situation where the transition frequencies behave as $\omega_{l,l-1} = l \omega$, we obtain a new expression with 
 \begin{eqnarray}
b_k(t_\infty) & = & \frac{1}{k!} \left( \frac{v_0 \tau_c}{\hbar}\right)^k
  D^+_{k-1} \cdots D^+_0  \prod_{l=1}^{k} 
  j_0 \left(l \omega \frac{\tau_c}{2} \right) , \nonumber
\end{eqnarray}
where $j_0$ is the first spherical Bessel function, $j_0(x) = \sin(x)/x$. 
If $|\omega \tau_c/2| \ll 1$, we will always have $|j_0 (l \omega \tau)| < |j_0 (\omega \tau)|$ that explains empirically 
why the approximations (\ref{eq:approx1}) leads systematically to an overestimation of the exact probabilities. In particular, we obtain an upper bound for the absolute value of $|b_k|$ that is given by:
\begin{eqnarray}
\frac{|b_k|^2}{|D^+_0|^2 \cdots |D^+_k|^2} & \le & \frac{1}{(k!)^4} \left(\frac{2v_0 }{\hbar \omega } \right)^{2k},  \nonumber 
\end{eqnarray}
This expression contains an extra $1/(k!)^2$ compared to the case where we assumed simply that all frequencies are the same. 

From this last remarks, we empirically assume that the proper combinatorial factors are those given by 
Eqs. (\ref{eq:approx1}) divided by $(k!)^2$, leading to:
\begin{align}
W_k   = & C^k_{\Omega_A - n_A} C^k_{n_B}  C^k_{n_A+k} C^k_{\Omega_B - n_B + k }, \label{eq:retain1} \\
W_{-k}  = &  C^k_{\Omega_B - n_B} C^k_{n_A}  C^k_{n_B+k} C^k_{\Omega_A - n_A + k } .  \label{eq:retain2} 
\end{align} 
This combinatorial factor improves significantly the description of the transfer probabilities and is the one retained in this work. 


\end{document}